\documentclass[aps,prd,twocolumn,amsmath,superscriptaddress,amssymb,showpacs,floatfix,nofootinbib,longbibliography]{revtex4-1}

\usepackage{graphicx}
\usepackage{dcolumn}
\usepackage{xcolor}
\usepackage{bm}
\usepackage{natbib}
\usepackage{calligra}
\usepackage[T1]{fontenc}
\usepackage{egothic}
\usepackage[T1]{fontenc}
\newfont{\rsfsten}{rsfs10 scaled 1200}
\newfont{\rsfsseven}{rsfs10 scaled 1200}
\newfont{\rsfsfive}{rsfs10 scaled 1200}
\usepackage{epsfig}
\usepackage{units}
\usepackage[utf8]{inputenc}
\usepackage{hyperref}

\usepackage[shortlabels]{enumitem}

\newcommand{\physrep}{{Physics~Reports}}
\newcommand{\araa}{{Annu.~Rev.~Astron.~Astrophys.}}
\newcommand{\aap}{{Astron.~Astrophys.}}
\newcommand{\apjl}{{Astrophys.~J.~Lett.}}

\newcommand{\aj}{{Astron.~J.}}
\newcommand{\mnras}{{Mon.~Not.~R.~Astron.~Soc.}}

\newcommand{\be}{\begin{equation}}
\newcommand{\ee}{\end{equation}}
\newcommand{\bea}{\begin{eqnarray}}
\newcommand{\eea}{\end{eqnarray}}

\newcommand{\GeV}{{\rm\; GeV}}

\newcommand{\PPC}{\texttt{PPPC4DMID}}

\def\lsim{\mathrel{\raise.3ex\hbox{$<$\kern-.75em\lower1ex\hbox{$\sim$}}}}
\def\gsim{\mathrel{\raise.3ex\hbox{$>$\kern-.75em\lower1ex\hbox{$\sim$}}}}

\begin{document}

\title{Revisiting GeV-scale annihilating dark matter with the \textit{AMS-02} positron fraction}

\author{Iason Krommydas}
\email{ik23@rice.edu, ORCID: orcid.org/0000-0001-7849-8863}
\affiliation{Physics Division, National Technical University of Athens, Zografou, Athens, 15780, Greece}
\affiliation{Department of Physics and Astronomy, Rice University, Houston, Texas, 77005, USA}
\author{Ilias Cholis}
\email{cholis@oakland.edu, ORCID: orcid.org/0000-0002-3805-6478}
\affiliation{Department of Physics, Oakland University, Rochester, Michigan, 48309, USA}
\date{\today}

\begin{abstract}
Antimatter cosmic-rays are used to probe new phenomena in physics, 
including dark matter annihilation. We use the 
cosmic-ray positron fraction spectrum by the Alpha Magnetic Spectrometer, 
to search for such an annihilation signal in the Galaxy.
We focus on dark matter with mass between 5 and 120 GeV, producing high-energy electrons and positrons.  
In these cosmic-ray energies the interplay of 
multiple astrophysical sources and phenomena, makes this search 
highly sensitive to the underlying astrophysical background assumptions. 
We use a vast public library of astrophysical models for the 
cosmic-ray positron fraction background, to derive robust upper limits on 
the dark matter's annihilation cross 
section for a number of annihilation channels. This library accounts for different types of cosmic-ray sources and uncertainties 
on their distribution in space and time. Also, it accounts for uncertainties on 
those sources' output, their injected 
into the interstellar medium cosmic-ray spectra and for uncertainties on cosmic-ray propagation. For any given dark matter particle mass and 
annihilation channel, upper limits on the annihilation cross section are given by bands that stretch a full order of magnitude in its value. 
Our work provides weaker limits compared to earlier results, 
that are however robust to all the relevant astrophysical uncertainties. 
Between 5 and 15 GeV, we find indications for a possible excess flux of cosmic-ray electrons and positrons. That excess is found for most, but not all of our astrophysical 
background parameter space, and its significance can vary appreciably. 
Further scrutiny is necessary to improve the understanding of these lower energy 
cosmic rays. Finally, we note that even if an excess signal is 
found in these energies, the current background uncertainties do not allow us to accurately deduce its underlying
particle properties. 

\end{abstract}

\maketitle

\section{Introduction}
\label{sec:introduction}

Dark matter has been observed in a variety of astrophysical systems through its
gravitational impact, in scales from as small as dwarf galaxies to as large as  
colliding galaxy clusters \cite{2007ApJ...670..313S, 2019ARA&A..57..375S, 2005ApJ...626L..85W, 2005AJ....129.2692W, Salucci:2018hqu, Catena:2009mf, Iocco:2011jz, Pato:2015dua, Bovy:2012tw, 2011ARA&A..49..409A, 2018PhR...733....1D, Clowe:2006eq, Brownstein:2007sr}. In addition, through detailed measurements 
of the cosmic microwave background (CMB), we know that dark matter accounts for about
$27 \%$ of the critical density in the universe, corresponding to about $85 \%$
of its matter 
\cite{Aubourg:2014yra, 2002ARA&A..40..643C, Cooke:2017cwo, Planck:2018vyg, Schoneberg:2019wmt}. 
Furthermore, accurate measurements probing Big Bang nucleosynthesis, the
evolution of structures in the universe, observations on the mass distribution of
different gravitationally collapsed structures and observations of the Layman-alpha
forest, set a strong preference for what is referred to as ``cold dark matter'' \cite{Iocco:2008va, Pospelov:2010hj, Cyburt:2015mya, Cooke:2017cwo, Cuoco:2003cu, 1991ApJ...379...52W, 2008ApJ...688..277T, Busca:2012bu, Aubourg:2014yra, 2012ARA&A..50..353K, 2010MNRAS.406.1759M, 2018PhR...733....1D, 1996ApJ...457L..51H, Garzilli:2019qki}. 
However, the nature of dark matter remains a puzzle, with its mass ranging from 
$10^{-22}$ eV to as large as $O(10) M_{\odot}$ \cite{Kobzarev:1966qya, Carr:1974nx, Meszaros:1974tb, 1975ApJ...201....1C, Weinberg:1977ma, Wilczek:1977pj, 1983PhLB..120..127P, Dine:1982ah, Abbott:1982af, Berezhiani:1995am, Svrcek:2006yi, Strassler:2006qa, Hooper:2008im, Arkani-Hamed:2008kxc, Hall:2009bx, Feng:2010gw, Graham:2015cka, Bertone:2016nfn, Hochberg:2014dra, Hochberg:2014kqa, Drewes:2016upu, Kahn:2016vjr, Bird:2016dcv, Carr:2016drx, Battaglieri:2017aum}.  

One class of dark matter candidates includes weakly interacting massive 
particles (WIMPs), that were thermally produced in the early universe 
through approximately electroweak scale interactions with Standard Model particles
and a mass very approximately of $O(10)$ GeV - $O(10)$ TeV \cite{Feng:2010gw, Steigman:2012nb, Bertone:2016nfn, Roszkowski:2017nbc, Griest:1989wd}. 
In this paper, we focus on the lower end of that mass range, probing dark matter 
with mass from 5 GeV and up to 120 GeV. For such dark matter particles 
we constrain the annihilation cross section they may have to leptons and to bottom 
quarks. Bottom quarks would be the prominent annihilation product for dark matter 
in that mass range, if dark matter couples to the Higgs boson \cite{Gunion:1989we}. 
These dark matter masses are interesting to search for, also because excesses in gamma-rays 
\cite{Goodenough:2009gk, Vitale:2009hr, Hooper:2010mq, Hooper:2011ti, Gordon:2013vta, Daylan:2014rsa, Calore:2014xka, Calore:2014nla, Abazajian:2014fta, Fermi-LAT:2015sau, DiMauro:2021raz, Cholis:2021rpp}
and in cosmic-ray antiprotons \cite{Cuoco:2016eej, Cui:2016ppb, Cholis:2019ejx, Cuoco:2019kuu} 
have been claimed to be compatible with WIMPs in that mass range. 
We use the most recent measurements of the cosmic-ray positron fraction, i.e. 
the ratio of the positron flux over the electron plus positron flux, versus 
those particle's energy; made by the Alpha Magnetic Spectrometer (\textit{AMS-02}) 
onboard the International Space Station \cite{AMS:2021nhj, AMS:2019rhg}. 

Over the last decades antimatter cosmic-ray measurements have been used to probe possible 
dark matter signals \cite{Bergstrom:1999jc, Hooper:2003ad, Profumo:2004ty, Bringmann:2006im, Pato:2010ih}. 
Such cosmic rays are produced from rare inelastic collisions between cosmic-ray nuclei with the 
interstellar medium 
(ISM) gas and are commonly referred to as secondary cosmic rays. Primary 
cosmic rays are instead accelerated in supernova remnant (SNR) environments. A 
hypothetical dark matter particle in the GeV-TeV mass scale annihilating (or decaying)
and producing among other byproducts antimatter cosmic rays, may give a 
detectable additional flux in measurements of such particles. This is the focus of this work.
Dark matter particles producing cosmic-ray positrons could cause a feature in the 
positron flux and positron fraction. The qualities of such a feature, depend on the dark matter 
particle's mass, annihilation cross section and channel, i.e. the fist generation of Standard Model 
particles produced from the annihilation event. Dark matter originated features 
may be as small as a localized in energy, to give a few $\%$ bump on the positron fraction, 
or as wide in energy and large in amplitude as the entire rising above 5 GeV positron 
fraction spectrum. 

Inversely, using the \textit{AMS-02} positron fraction's relatively 
smooth spectrum, one can set upper limits on the annihilation 
cross section of dark matter particles. That is done for a range of masses and a variety of annihilation 
channels \cite{Bergstrom:2013jra, Ibarra:2013zia, John:2021ugy}. 
This is the main aim of this paper. 

The origin of the rising above 5 GeV positron fraction 
spectrum that was first measured by the Payload for Antimatter Matter Exploration 
and Light-nuclei Astrophysics (\textit{PAMELA}) satellite \cite{2009Natur.458..607A, PAMELA:2013vxg}, 
then confirmed by the \textit{Fermi}-LAT \cite{2012PhRvL.108a1103A} and further measured with 
an unprecedented accuracy by \textit{AMS-02} \cite{AMS:2014bun, AMS:2019iwo, AMS:2021nhj}, 
has been a subject of great interest. 
One explanation for the additional positron flux, is relatively close-by ``young'' 
and ``middle-aged'' Milky Way pulsars that during their pulsar wind nebula (PWN) 
phase converted an appreciable fraction ($O(0.01)-O(0.1)$) of their rotational 
energy into high-energy cosmic-ray electrons ($e^{-}$) and positrons ($e^{+}$)  
\cite{1987ICRC....2...92H, 1995PhRvD..52.3265A, 1995A&A...294L..41A, Hooper:2008kg, Yuksel:2008rf, Profumo:2008ms, Malyshev:2009tw, Kawanaka:2009dk, Grasso:2009ma, 2010MNRAS.406L..25H, Linden:2013mqa, Cholis:2013psa, Yuan:2013eja, Yin:2013vaa, Cholis:2018izy, Evoli:2020szd, Manconi:2021xom, Orusa:2021tts, Cholis:2021kqk}. 
Another explanation is Milky Way SNRs, that in their first $O(10)$ kyr
produced and accelerated secondary cosmic rays including positrons
\cite{Blasi:2009hv, Mertsch:2009ph, Ahlers:2009ae, Blasi:2009bd, Kawanaka:2010uj, Fujita:2009wk, DiMauro:2014iia, Kohri:2015mga, Mertsch:2018bqd} 
(see however \cite{Cholis:2013lwa, Mertsch:2014poa, Cholis:2017qlb, Tomassetti:2017izg}). 
Furthermore, detailed modifications on the distribution of cosmic-ray 
sources and the propagation of cosmic rays through the ISM \cite{Gaggero:2013rya, Cowsik:2013woa, Gaggero:2013nfa} and 
annihilating or decaying dark matter models have been explored to explain the 
positron fraction measurement \cite{Bergstrom:2008gr, Cirelli:2008jk, Cholis:2008hb, Cirelli:2008pk, Nelson:2008hj, ArkaniHamed:2008qn, Cholis:2008qq, Cholis:2008wq, Harnik:2008uu, Fox:2008kb, Pospelov:2008jd, MarchRussell:2008tu, Chang:2011xn, Cholis:2013psa, Dienes:2013xff, Finkbeiner:2007kk, Kopp:2013eka, Dev:2013hka, Klasen:2015uma, Yuan:2018rys, Sun:2020dla}. 
We assume in this work that the overall rise of the positron  fraction, shown 
with its \textit{AMS-02} measurement in Fig.~\ref{fig:pf}, is not caused by 
dark matter, but instead from a more conventional source; a 
population of Milky Way pulsars. 

Pulsars are localized sources of cosmic-ray electrons and positrons. 
Due to their rapid spin-down, pulsars convert their initial rotational energy 
into cosmic-ray $e^{\pm}$ and subsequently release those $e^{\pm}$ into the 
ISM in a comparatively short amount of time\footnote{The time required for most cosmic-ray $e^{\pm}$ 
produced around the PWN environments to be released into the ISM, is at 
least an order of magnitude smaller than the propagation time required for 
these cosmic rays to reach our detectors \cite{Malyshev:2009tw}. The only 
exception would be a very close ($O(10)$ pc) pulsar 
(see however \cite{Evoli:2018aza}).}.
That makes pulsars cosmic-ray $e^{\pm}$ sources approximately 
localized both in space and time. High-energy $e^{\pm}$ lose 
rapidly their energy through synchrotron radiation and inverse Compton scattering 
as they interact with the ISM and before reaching us.
That results in an upper
energy cut-off, on the $e^{\pm}$ spectra from individual pulsars
\cite{Profumo:2008ms, Malyshev:2009tw, Grasso:2009ma}. In turn, a population 
of pulsars that could collectively explain the rising positron 
fraction spectrum, could also give spectral features at the higher energies where
the number of contributing pulsars is reduced to the point of individual
sources dominating narrow parts of that spectrum \cite{Malyshev:2009tw, Grasso:2009ma, Cholis:2018izy, Cholis:2021kqk}. Such features can then
be searched for as in \cite{Cholis:2017ccs}. Similar arguments can be made 
for PWNe. However, their expected higher energy cut-offs are less sharp by 
comparison \cite{Mertsch:2018bqd}. We use modeled populations of Milky Way 
pulsars produced in our earlier work of \cite{Cholis:2021kqk}. In 
Ref.~\cite{Cholis:2021kqk}, a library of publicly available pulsar population 
models was created that is in agreement with the cosmic-ray $e^{\pm}$ flux 
spectral measurements from \textit{AMS-02} \cite{AMS:2019rhg, AMS:2021nhj}, 
the CALorimetric Electron Telescope (\textit{CALET}) \cite{Adriani:2018ktz} 
and the DArk Matter Particle Explorer (\textit{DAMPE}) telescope \cite{DAMPE:2017fbg}, as
well as the \textit{AMS-02} positron fraction spectrum \cite{AMS:2019iwo}.
As the pulsar's contribution to the positron fraction spectrum is not 
perfectly smooth and with uncertainties, we use a library of
models instead of just one generic parameterization. As we will show, we derive 
more conservative and more realistic limits on the dark matter annihilation 
cross section. 

In Section~\ref{sec:methodology}, we discuss the general methodology of our
approach, including the observations that we use, 
the astrophysical background modeling of the positron fraction and the 
statistical treatment followed in fitting the data. We also create mock
positron fraction data to answer the question on the robustness of the positron 
fraction measurement as a means to study the particle properties of dark matter. Then in 
Section~\ref{sec:results}, we present the results of searching for a possible
dark matter signal in the positron fraction. We find that the limits on the
annihilation cross section are not well defined. That is due to the underlying 
astrophysical background uncertainties. The annihilation cross section limits 
have a width that is at least one order of magnitude in the mas range of 5 to 120 
GeV that we study. In addition, we find indications for a possible excess of 
5-15 GeV in cosmic-ray energy $e^{\pm}$. That excess while compatible 
with a WIMP-scale dark matter signal, has a significance that varies with the 
astrophysical background modeling and is not claimed to be a robust one. 
Further scrutiny will be required as cosmic-ray physics in that energy range improve with 
future observations. Moreover, in Section~\ref{sec:results}, we perform our
mock positron fraction analysis. We find that if dark matter contributes 
to the positron fraction spectrum at the few percent level within an range spanning
several \textit{AMS-02} energy bins, such an excess signal can not be absorbed 
by the astrophysical background uncertainties. However, identifying the exact
particle properties of the dark matter particle responsible for that excess is a
a more model-dependent inquiry. Finally, in Section~\ref{sec:conclusions},
we give our conclusions and discuss connections to other types of dark matter
searches as well as future prospects. 

\section{Methodology}
\label{sec:methodology}

In this section, we describe the energy range of the \textit{AMS-02} positron
fraction ($e^{+}/(e^{+} + e^{-})$) measurement used in this analysis.  
We also explain how we construct our background astrophysical models, which are 
fitted to the \textit{AMS-02} positron fraction. 
We then describe the statistical analysis performed to set upper limits on dark 
matter particles annihilating, giving a contribution to the positron fraction.
Finally, we construct positron fraction mock data based on the \textit{AMS-02} 
sensitivity to test whether a dark matter signal would be detectable; and how 
accurately we would be able to determine the dark matter mass, annihilation channel 
and cross section by our analysis.

We use astrophysical realizations created within Ref.~\cite{Cholis:2021kqk}, as a 
base to construct our background models for the positron fraction.
We take the $e^-$ and $e^+$ fluxes calculated from these realizations and add a 
dark matter contribution. Using these fluxes we perform fits, where we search 
for a potential dark matter component to the positron fraction and compute the 
95\% confidence level upper limits on the dark matter annihilation cross section 
as a function of mass. These fits are performed using a library of astrophysical/background realizations.   

\subsection{Cosmic-ray data}
\label{subsec:data}

We use the recently published \textit{AMS-02} positron fraction measurement from \cite{AMS:2021nhj,AMS:2019rhg} taken between May 2011 and November 2017. 
In Ref.~\cite{Cholis:2021kqk}, we found that the positron fraction spectrum sets stronger constraints on sources of cosmic-ray positrons, compared to the cosmic-ray positron flux spectrum. This is due to its smaller errors. Some systematic errors cancel when calculating cosmic-ray fractions versus cosmic-ray fluxes.
We ignore the positron fraction measurement below 5 GeV, as that energy range is 
strongly affected by solar modulation and any dark matter annihilation signal from 
an approximately thermal relic would be hidden within the solar modulation modeling 
uncertainties. 
Given that there is no publicly released covariance matrix by the \textit{AMS-02} 
collaboration on that measurement, we treat the different energy bins as 
uncorrelated and add the systematic and statistical errors in quadrature. 

\subsection{Modeling the background to the dark matter contribution on the positron fraction}
\label{subsec:modeling}

In this work, as signal we refer to a potential annihilating dark matter contribution on the 
positron fraction spectrum. As background we refer to all other astrophysical sources contributing to the positron fraction.
Our modeling of the astrophysical background is based on Ref.~\cite{Cholis:2021kqk}.
These astrophysical realizations contain $e^-$ fluxes from primary sources 
i.e. supernova remnants, secondary $e^{\pm}$ produced from inelastic 
collisions of primary cosmic ray nuclei with the ISM gas and $e^{\pm}$ 
from Milky Way pulsars.
The main goal of Ref.~\cite{Cholis:2021kqk}, was to study the properties of 
Milky Way pulsars. Thus, a large number of astrophysical realizations was created. 
Those realizations accounted for a sequence of astrophysical uncertainties, as the
stochastic nature of the neutron stars' birth in time and location, 
the stochasticity in the initial spin-down power of pulsars and their 
subsequent time evolution. Also Ref.~\cite{Cholis:2021kqk}, studied the 
fraction of pulsar spin-down power into cosmic-ray $e^{\pm}$ and how these 
injected cosmic rays propagate in the ISM and the Heliosphere. 

In this work we start with the astrophysical/background realizations from Ref.~\cite{Cholis:2021kqk} that were shown to be in good agreement with the \textit{AMS-02} positron fraction \cite{AMS:2021nhj,AMS:2019rhg}, the $e^+$ flux \cite{AMS:2021nhj,AMS:2019rhg}, the total $e^++e^-$ flux \cite{AMS:2021nhj,AMS:2019iwo}, and also the total  $e^++e^-$ fluxes from \textit{DAMPE} \cite{Ambrosi:2017wek} and \textit{CALET} \cite{Adriani:2018ktz}.
The quality of the fit is heavily impacted by the lowest energies of the positron fraction where the errors are the smallest.
Adding a dark matter $e^{\pm}$ flux component that contributes at these low energies can drastically affect the quality of the fit.
Thus, we include in our analysis astrophysical/background realizations from Ref.~\cite{Cholis:2021kqk} that have a $\chi^2/n_{\mathrm{dof}} < 2.2$ in the positron fraction. This results in a total of 1020 astrophysical/background simulations, to account for all the background uncertainties. 
Some of those realizations in combination with a dark matter component may end up
giving a much better quality of fit to the \textit{AMS-02} data and can explain 
the $e^{\pm}$ observations at energies where there is no contribution from dark matter. 

For the dark matter contribution, we assume a local dark matter density of $0.4\ \rm GeV/cm^3$ \cite{Bovy:2012tw,Salucci:2010qr,Catena:2009mf,Pato:2015dua}, set at 8.5 kiloparsec (kpc) from the galactic center. 
We take the dark matter halo in the Galaxy to follow an Navarro-Frenk-White (NFW) profile \cite{Navarro:1995iw}, with a characteristic radius of 20 kpc.
We consider four simplified dark matter annihilation channels. 
These are:  $\chi\chi \rightarrow e^+ e^-$, $\chi\chi \rightarrow \mu^+ \mu^-$, $\chi\chi \rightarrow \tau^+ \tau^-$ and $\chi\chi \rightarrow b\bar{b}$.
The annihilation cross section is set to be free in our analysis. 
We focus on low dark matter masses $m_{\chi}$ between 5 and 50 GeV for the $e^+ e^-$ and the $\mu^+ \mu^-$ channels, between 5 and 80 GeV for the $\tau^+ \tau^-$ channel and between 10 and 120 GeV for the $b\bar{b}$ one.

We calculate the injected $e^{\pm}$ production spectra from these dark matter annihilations using \PPC \cite{Cirelli:2010xx} and calculate the final $e^-$ and $e^+$ spectra at the location of the Sun using
\path{GALPROP} v54 \cite{galpropwebsite, galprop}.
The dark matter $e^{\pm}$ spectra are propagated through the ISM using the same 12 alternative propagation models as those defined in Table II of Ref.~\cite{Cholis:2021kqk}. 
Every time that we test for a potential dark matter signal in the \textit{AMS-02} data, we make sure that the hypothetical dark matter $e^{\pm}$ flux, is evaluated under the same propagation conditions as its relevant astrophysical background. 
The 12 ISM models account for different choices on the thickness of the zone 
within which cosmic rays diffuse before escaping the Milky Way, how that diffusion
depends on the cosmic-ray energy and finally for the energy losses of the cosmic-ray $e^{\pm}$ within the local volume of the Milky Way. 
This combination of ISM models encompasses the relevant astrophysical uncertainties within $O(\textrm{kpc})$ from the Sun \cite{Trotta:2010mx, Pato:2010ih, Cholis:2021rpp}. 
For more details we refer the reader to Section II.E of Ref.~\cite{Cholis:2021kqk}.
In each astrophysical background, we add a dark matter contribution by choosing a specific annihilation channel and a specific mass and construct our final astrophysical+dark matter model.
Given the different choices for the particle dark matter properties, we simulate 64 different combinations of annihilation channel and mass for each of the 1020 astrophysical backgrounds (65280 fits in total).
The annihilation cross-section is left as a free parameter to be set by the fit to the data.
These final ISM $e^-$ and $e^+$ spectra include the contribution of primary cosmic rays, secondary cosmic rays, cosmic rays from pulsars and from dark matter annihilations.
We also propagate each of the ISM cosmic-ray spectra to the location of the Earth 
and account for solar modulation. That is done following the prescription of 
\cite{Cholis:2015gna}, where the modeling of the time, charge and energy-dependence 
of solar modulation is accounted for by two fitting parameters, set within a range 
suggested by \cite{Cholis:2020tpi, Cholis:2022rwf}. That same procedure was 
followed in Ref.~\cite{Cholis:2021kqk}. The associated Bartels' Rotation numbers
-relevant for the modeling of solar modulation effects- for the data-taking era are 
2426-2514.

\subsection{Statistical analysis}
\label{subsec:analysis}

When we fit the astrophysical/background models to the \textit{AMS-02} positron 
fraction we have seven parameters. These account for the cosmic-ray primary $e^-$ 
flux, the secondary $e^{\pm}$ flux and the pulsar $e^{\pm}$ flux normalizations.
We include two parameters to allow for a spectral softening/hardening of the 
cosmic-ray primary and secondary spectra; and two more for the solar modulation 
modeling. 
Once adding the dark matter component we have an additional (eighth) parameter, 
that is directly proportional to the fitted annihilation cross section. 
For a given astrophysical background, once adding a potential dark matter component 
in the fitting procedure, we allowed the other seven parameters to be free within 
$50\%$ of their best-fit value achieved in the background only fit.
In Appendix~\ref{app:table} we give the full parameter space tested in our 
astrophysical $\&$ dark matter models and its subsequent minimization procedure.

We perform a $\chi^2$ minimization; and use a combination of \path{SciPy}'s
\cite{2020SciPy-NMeth} \path{least_squares} routine from the \path{optimize} module 
and \path{iminuit} \cite{iminuit,James:1975dr}.
We found that the fastest minimization is achieved by performing a few minimization 
steps with the \path{least_squares} routine with high tolerance and finishing the 
minimization with \path{iminuit}.

\begin{figure}[!t]
    \centering
    \includegraphics[width=1\linewidth]{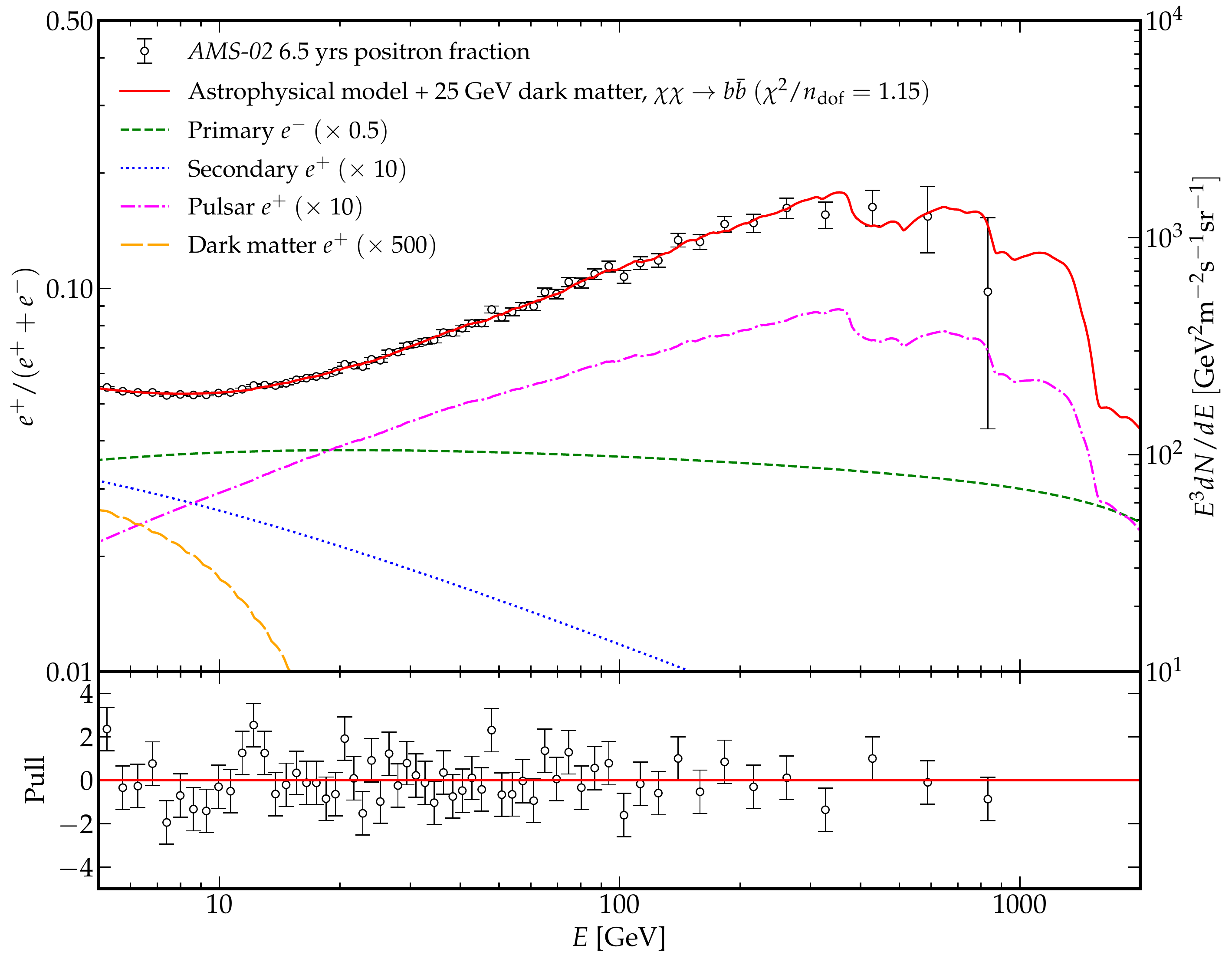}
    \caption{The fit of a pulsar model from Ref.~\cite{Cholis:2021kqk}, to the \textit{AMS-02} positron fraction after including the contribution from 25 GeV dark matter that annihilates to $b\bar{b}$.
    On the right y-axis in units of $E^{3}dN/dE$ where $dN/dE$ is the differential cosmic-ray flux, we show the solar modulated contribution from primary $e^-$, secondary $e^+$, pulsar $e^+$, and $e^+$ fluxes originating from dark matter annihilation scaled by some appropriate arbitrary factors from their best fit normalization to make them well-visible.
    The secondary, pulsar and dark matter $e^-$ fluxes are not shown since they are only slightly different due to solar modulation.
    We also show the pull ($(\mathrm{data-model})/\sigma_{\mathrm{data}}$) distribution of the fit at the bottom.}
    \label{fig:pf}
\end{figure}

In Fig.~\ref{fig:pf}, we show the fit to the positron fraction for one of our 
background models with a dark matter component included. For the dark matter we 
have taken, $m_{\chi}=25\ \rm GeV$ and the annihilation channel to be $\chi\chi \rightarrow b\bar{b}$.
One can see all the relevant contributions from primary $e^-$, secondary $e^+$, 
pulsar $e^+$, and $e^+$ originating from dark matter.
The background only hypothesis gave for this model a  $\chi^2_{\mathrm{DM=0}}/n_{\mathrm{dof}}=2.11$; while after including dark matter 
we got a $\chi^2_{\mathrm{DM}}/n_{\mathrm{dof}}=1.15$. 
$n_{\mathrm{dof}}$ is the number of degrees of freedom.
The pull of the fit, which is $(\mathrm{data-model})/\sigma_{\mathrm{data}}$ is 
also shown at the bottom of the figure.

The dark matter mass range that we study in this work, contributes at the lower
energies and has no effect on the higher energies where the spectrum is dominated 
by the local pulsar population. Also, other than the $\chi\chi \rightarrow e^+ e^-$ 
annihilation channel, the dark matter component cannot produce sharp peaks in the 
positron fraction that could explain features like the one we identified in 
Ref.~\cite{Cholis:2021kqk} and studied in Ref.~\cite{Cholis:2022ajr} at $\sim 12\ \rm GeV$. 
We find that statement to be true for every astrophysical/background model. 

In order to derive upper limits on the dark matter annihilation cross section, 
presented as $\langle \sigma v \rangle$, we use a likelihood ratio test.
The null hypothesis is that there is no dark matter contribution and we just have 
the astrophysical background, i.e. seven fitting parameters with $\langle \sigma v \rangle =0$. 
The alternative hypothesis is that there is some contribution from dark matter 
annihilation with annihilation cross section (times velocity, thermally averaged) $\langle \sigma v \rangle$, i.e. eight fitting parameters.
We rely on Wilks' theorem \cite{10.1214/aoms/1177732360}, and use the statistic 
$LR=-2\log{\Lambda}$(=$\chi^2$ difference), with $\Lambda$ the likelihood ratio of 
the null (background only) hypothesis over the alternative dark matter + background 
hypothesis. This is distributed according to a $\chi^2_{\nu}$-distribution with 
$\nu$ degrees of freedom, where $\nu$ is the difference of fitting parameters 
between the two hypotheses models. In our case, we have $\nu=1$.
However, that would give a naive estimate of the p-value since the null hypothesis
corresponds to the case $\langle \sigma v \rangle=0$, i.e. it lies on a boundary of our parameter 
space. This problem can be overcome by using Chernoff’s theorem 
\cite{10.1214/aoms/1177728725}. The $LR$ follows a $\frac{1}{2} \delta(x) + \frac{1}{2}\chi^2$ distribution (half chi-square distribution) with one degree of 
freedom \cite{Conrad:2014nna}.
This means that the p-value is reduced by half compared to the naive estimate.

Following the standard convention in the literature, we can deduce 95\% upper 
limits on $\langle \sigma v \rangle$ for each astrophysical background at a fixed annihilation 
channel and dark matter mass.
This is done by scanning over $\langle \sigma v \rangle$, 
computing the $\chi^2$ profile and finding at which value of 
$\langle \sigma v \rangle$ we have $\chi^2_{\mathrm{DM}}=\chi^2_{\mathrm{DM=0}}+2.71$. This
corresponds to the 95\% upper limit of a half chi-square distribution with 
one degree of freedom. Because we have multiple masses, we essentially have a 
2D grid of masses and cross sections where we compute the $\chi^2$ profile and draw 
the contour where the $\chi^2_{\mathrm{DM}}$ increased by 2.71 from 
$\chi^2_{\mathrm{DM=0}}$. At each point of the grid the rest of the background 
nuisance parameters are optimized such that the $\chi^2$ is minimum. This contour 
is the 95\% upper limit on the dark matter annihilation cross section as a 
function of the mass.
This can be done for each annihilation channel and for each pulsar background,
resulting in each background giving a different upper limit. We report 
the combination of those upper limits.

\subsection{Mock data}
\label{subsec:mockdata}

In this paper, we produce mock data of the \textit{AMS-02} positron fraction.
We do that to test whether a dark matter contribution in the positron fraction 
would be detectable and with its properties (mass, annihilation cross section 
and channel) correctly identified. 
We produce these mock data by taking existing backgrounds and adding a flux component from dark matter of specific mass, annihilation channel and cross section.
We then calculate the positron fraction spectra that the \textit{AMS-02} would observe.
These mock spectra include only statistical errors.
We treat these mock spectra as we treated the \textit{AMS-02} measurement 
and scan them with our background+dark matter models to see if we can recover 
the original mass, annihilation channel and cross section.
By keeping only the statistical errors we are optimistic on the ability of the 
positron fraction measurement to help us probe the properties of a dark matter 
signal.

For the mock positron fraction spectra we use two annihilation channels: 
$\chi\chi \rightarrow \mu^+ \mu^-$ and $\chi\chi \rightarrow \tau^+ \tau^-$; and test four dark matter mass and annihilation cross section combinations.
For the $\chi\chi \rightarrow \mu^+ \mu^-$ channel we have,
\begin{enumerate}[(a)]
\item{$m_{\chi}=15\ \GeV$ and $\langle \sigma v \rangle = 2\times 10^{-26}\ \mathrm{cm^3 s^{-1}}$},
\item{$m_{\chi}=15\ \GeV$ and $\langle \sigma v \rangle = 5\times 10^{-27}\ \mathrm{cm^3 s^{-1}}$},
\item{$m_{\chi}=30\ \GeV$ and $\langle \sigma v \rangle = 2\times 10^{-27}\ \mathrm{cm^3 s^{-1}}$},
\item{$m_{\chi}=30\ \GeV$ and $\langle \sigma v \rangle = 5\times 10^{-28}\ \mathrm{cm^3 s^{-1}}$}.
\end{enumerate}
For the $\chi\chi \rightarrow \tau^+ \tau^-$ channel we have,
\begin{enumerate}[(a)]
\item{$m_{\chi}=15\ \GeV$ and $\langle \sigma v \rangle = 1\times 10^{-25}\ \mathrm{cm^3 s^{-1}}$},
\item{$m_{\chi}=15\ \GeV$ and $\langle \sigma v \rangle = 2\times 10^{-26}\ \mathrm{cm^3 s^{-1}}$},
\item{$m_{\chi}=30\ \GeV$ and $\langle \sigma v \rangle = 2.5\times 10^{-26}\ \mathrm{cm^3 s^{-1}}$},
\item{$m_{\chi}=30\ \GeV$ and $\langle \sigma v \rangle = 5\times 10^{-27}\ \mathrm{cm^3 s^{-1}}$}.
\end{enumerate}
We use two astrophysical backgrounds to create these mock data, one for each channel.
These two backgrounds are in agreement with the \textit{AMS-02} $e^{+}$ flux, 
positron fraction and total $e^{+} + e^{-}$ flux measurements, and also with 
the \textit{DAMPE} and \textit{CALET} total $e^{+} + e^{-}$ flux measurements.

\begin{figure}[!t]
    \centering
    \includegraphics[width=1\linewidth]{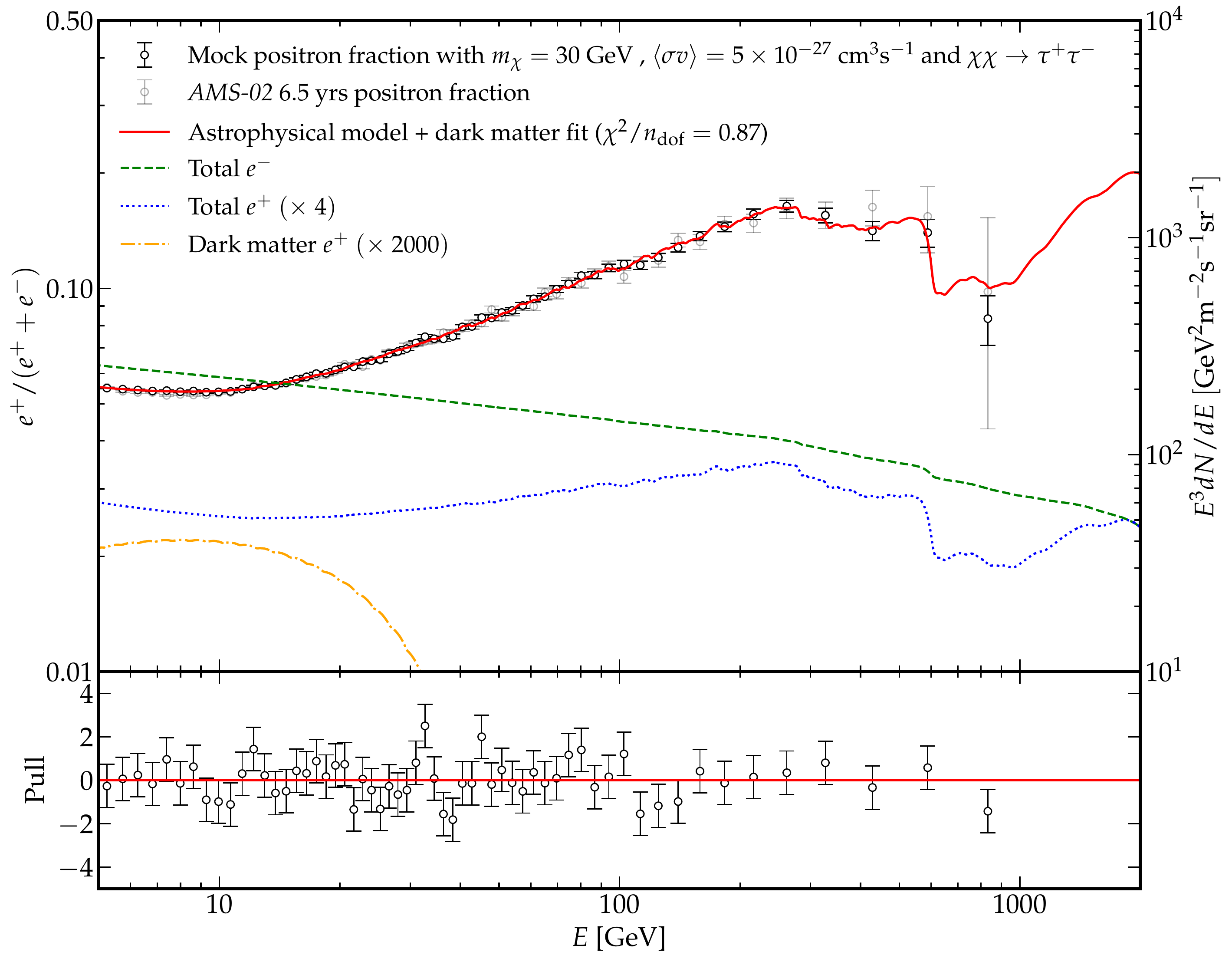}
    \caption{A mock positron fraction created by the combination of an astrophysical background with a dark matter signal. We chose an astrophysical background that without the dark matter contribution gave a good fit to the positron fraction ($\chi^2/n_{\mathrm{dof}}=1.14$). We add the contribution from a 30 GeV dark matter particle that annihilates to $\tau^+\tau^-$ with a cross section of $\langle \sigma v \rangle = 5\times 10^{-27}\ \mathrm{cm^3 s^{-1}}$.
    We show an astrophysical background+dark matter model fit to the mock positron fraction and also the total $e^-$, $e^+$ and $e^+$ originating from dark matter fluxes within that model in the green dashed, blue dotted and orange dash-dotted lines with the units provided by the right y-axis. At the bottom, 
    we show the pull distribution of the fit.
    We also show with faint grey the real \textit{AMS-02} positron fraction measurements.}
    \label{fig:mockpf}
\end{figure}

In Fig.~\ref{fig:mockpf}, we show an example of such a mock positron fraction 
where there is a contribution from 30 GeV dark matter particles annihilating to 
$\tau^+\tau^-$ with an cross section of $\langle \sigma v \rangle = 5\times 10^{-27}\ \mathrm{cm^3 s^{-1}}$.
We also show a fit to that mock positron fraction by one of our models and the 
original \textit{AMS-02} data with a fainter color for comparison to the mock ones.
The errors of the mock positron fraction are much smaller as we only consider statistical uncertainties.

Once a mock positron fraction spectrum that includes a dark matter component is created, we test our ability to deduce the particle dark matter properties.   
We scan the mock positron fraction spectra, by testing the combination of 100 backgrounds (including the backgrounds used to create them) with all the 64 combinations for the dark matter annihilation channel and mass used.

\section{Results}
\label{sec:results}

\subsection{Upper limits on the dark matter
annihilation cross-section}
\label{subsec:limits}

\begin{figure*}
\begin{centering}
\hspace{-0.1cm}
\includegraphics[width=3.50in,angle=0]{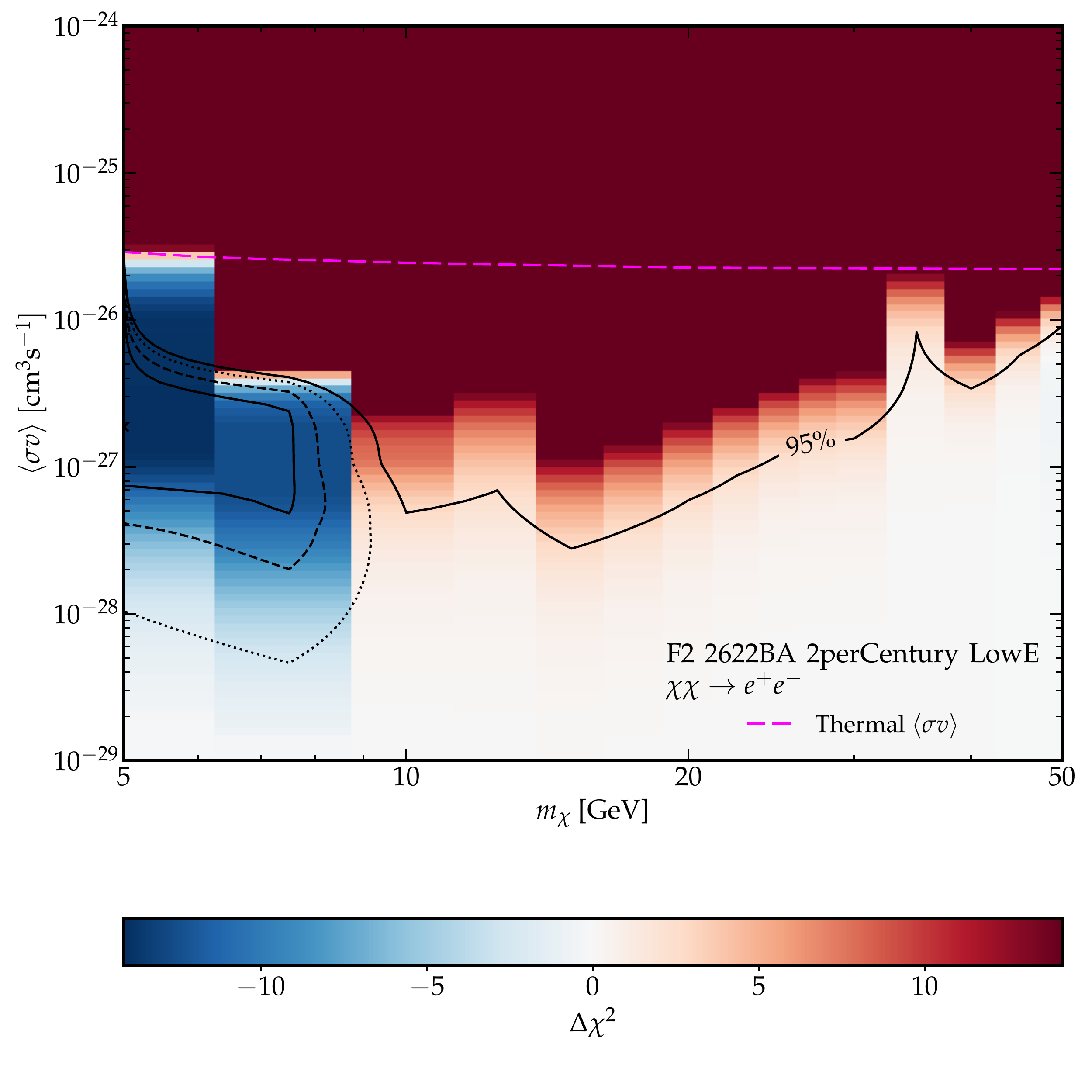}
\includegraphics[width=3.50in,angle=0]{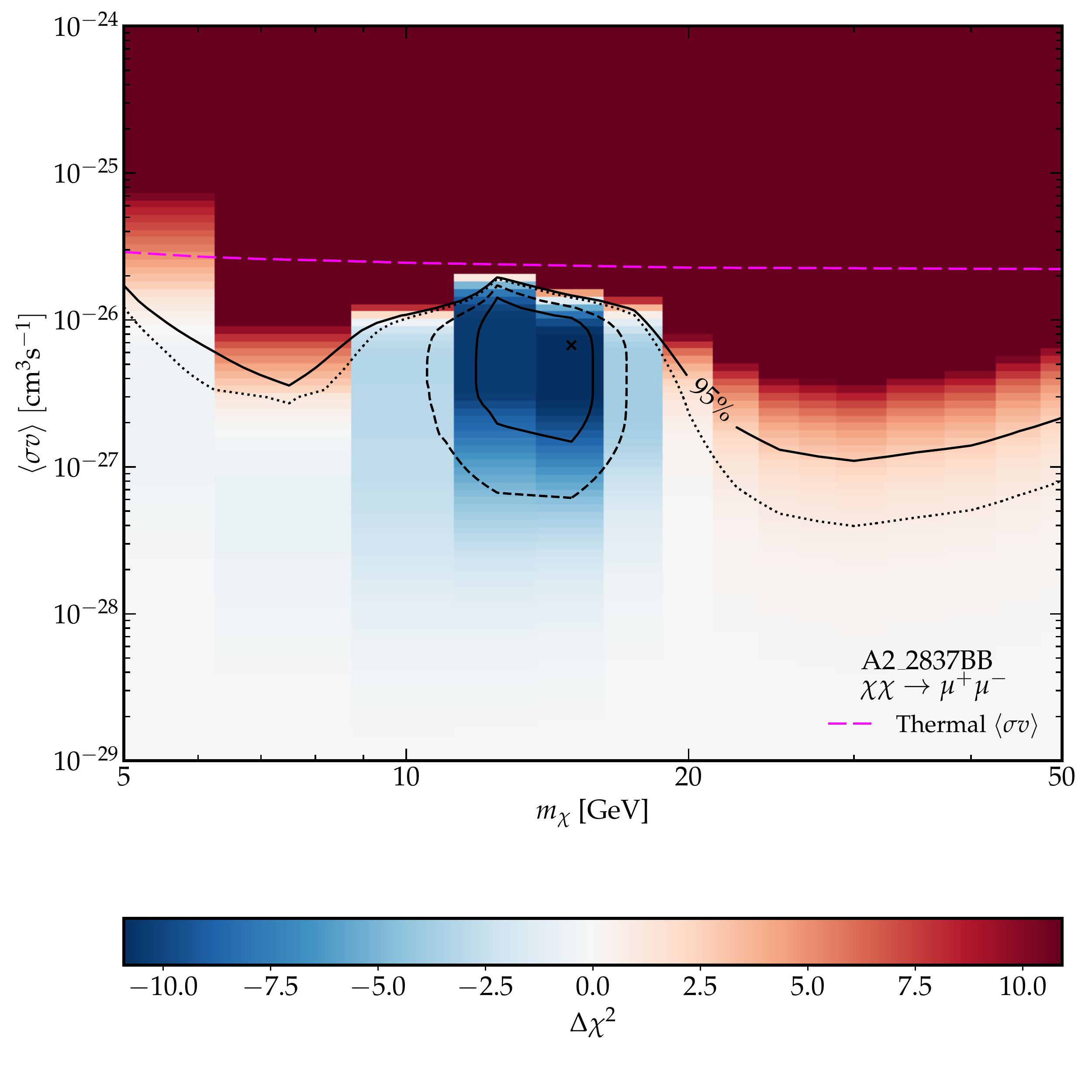}
\end{centering}
\vspace{-0.5cm}
\vspace{-0.0cm}

\begin{centering}
\hspace{-0.1cm}
\includegraphics[width=3.50in,angle=0]{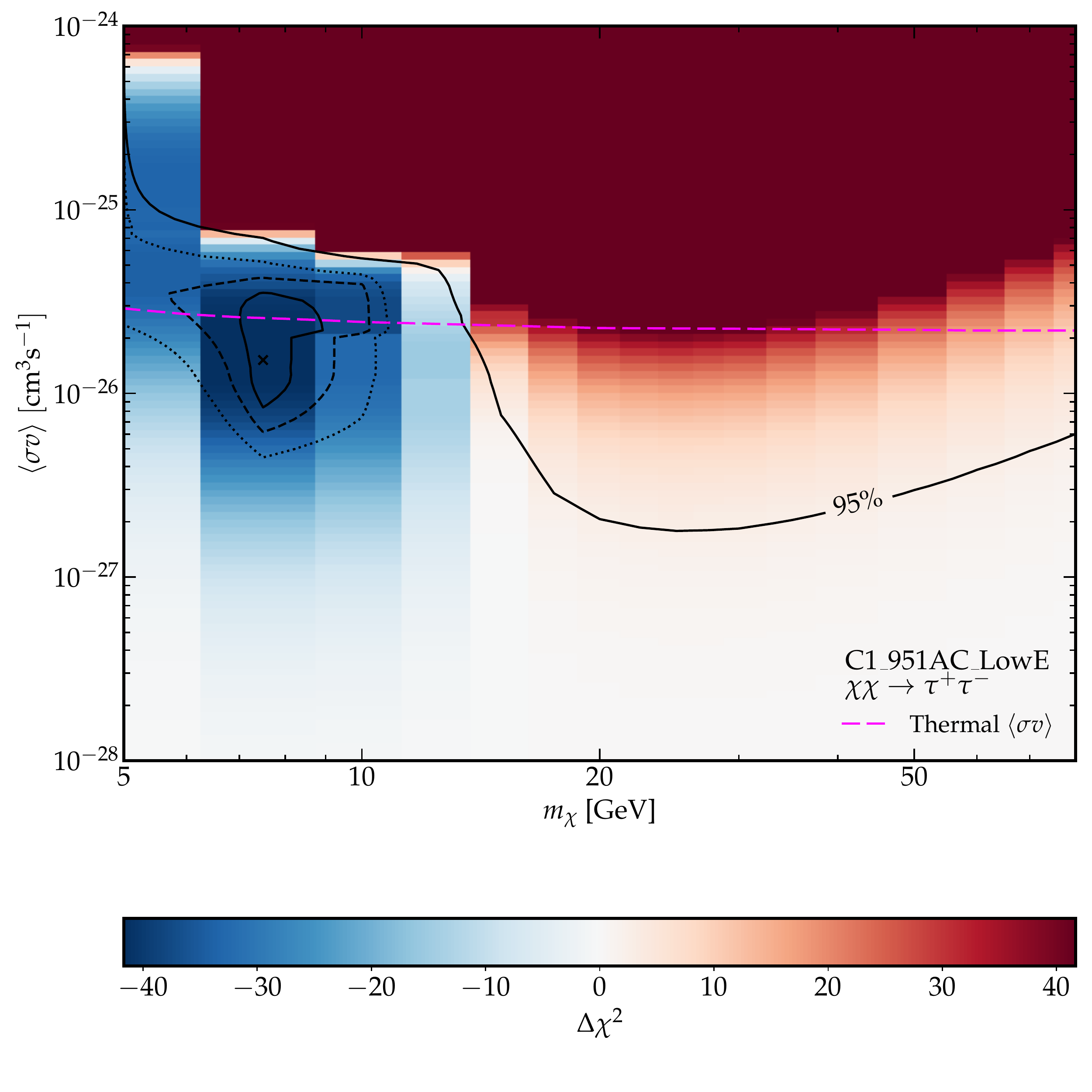}
\includegraphics[width=3.50in,angle=0]{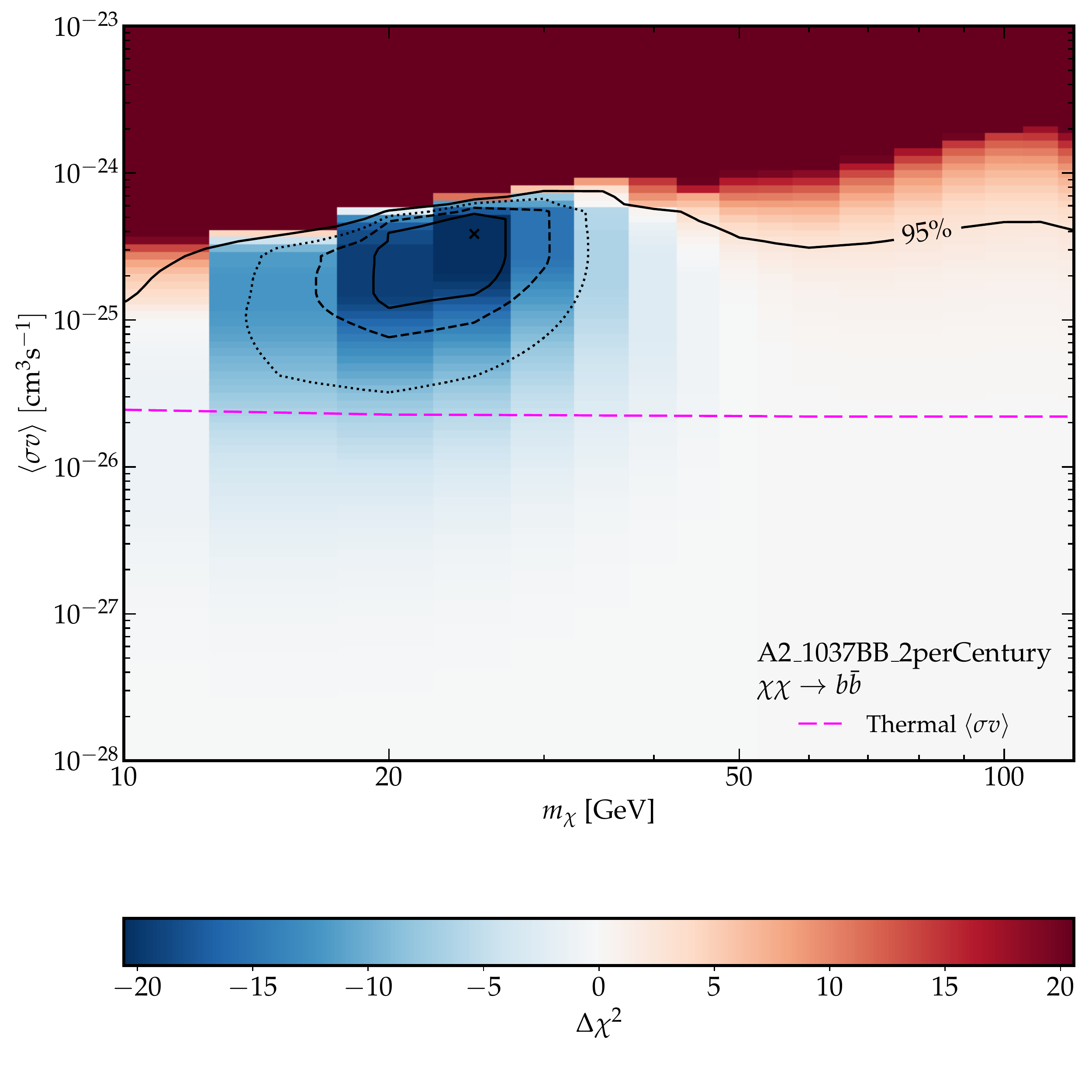}
\end{centering}
\vspace{-0.5cm}
\caption{The $\chi^2$ profiles presented as heatmaps for four different astrophysical backgrounds with the addition of dark matter annihilating to $e^{+} e^{-}$ (upper left), $\mu^+ \mu^-$ (upper right), $\tau^+ \tau^-$ (lower left) and $b \bar{b}$ (lower right).
In each heatmap we mark the best fit point with an "x" and draw the $1\sigma$ (solid), $2\sigma$ (dashed) and $3\sigma$ (dotted) contours around the best fit point. In the colorbars we show the $\Delta \chi^2$ values from fitting just the astrophysical background without the addition of annihilating dark matter.
We also show the 95\% upper limit lines on $\langle \sigma v \rangle$ for each one of these four backgrounds as constructed in subsection \ref{subsec:analysis}.
Finally, the dashed magenta lines are the expected $\langle \sigma v \rangle$ for a thermal relic from Ref.~\cite{Steigman:2012nb}.
The left two plots are with low-energy extrapolated backgrounds while the right two ignore the low-energy extrapolation (see text for details and also Ref.~\cite{Cholis:2021kqk}).}
\vspace{-0.0cm}
\label{fig:heatmaps}
\end{figure*}

For a given astrophysical background and after fixing the annihilation channel 
and dark matter mass $m_{\chi}$, we can calculate a unique limit of the dark 
matter annihilation cross section. 
In this work we show results for the annihilation cross section times the relative 
dark matter particles' velocity $\langle \sigma v \rangle$ (thermally averaged).
In Fig.~\ref{fig:heatmaps}, we show for the $m_{\chi}$ versus $\langle \sigma v \rangle$ parameter 
space the quality of $\chi^{2}$ fits as ``heatmaps'', for the four annihilation 
channels that we study. 
In each case these are evaluated by fixing the astrophysical background to be one 
out of the 1020 that we use as a basis to account for astrophysical modeling 
uncertainties. As we have explained in section~\ref{subsec:analysis}, we allow for 
eight fitting parameters. In these heatmaps the $\Delta \chi^{2}$ is evaluated from 
the case of no dark matter annihilation (i.e. $\langle \sigma v \rangle =0$). 
Points with negative $\Delta \chi^{2}$, (given in different shades of blue) 
represent dark matter assumptions for which we get a better fit to the positron 
fraction than without a dark matter contribution.
Instead, points with positive $\Delta \chi^{2}$, (given in different shades of red) 
represent dark matter assumptions that are statistically excluded by the positron 
fraction data. Our discretized parameter space shows the grid points that we used 
to probe the $m_{\chi}$ versus $\langle \sigma v \rangle$ parameter space. 

As can be seen from Fig.~\ref{fig:heatmaps}, for each channel and for the 
specific astrophysical backgrounds shown, there is a strong preference for a dark 
matter contribution to the positron fraction. This is shown by the dark blue ranges 
around which we evaluate the best fit $m_{\chi}$ and $\langle \sigma v \rangle$ parameter point and 
the 1, 2 and 3$\sigma$ signal significance contours. We also show the 95$\%$ upper 
limit on $\langle \sigma v \rangle$ as a function of $m_{\chi}$ (in the solid black lines)\footnote{These upper upper limit lines are only exact at the centers of our mass "bins". It shouldn't be worrying that the line goes over areas of negative $\Delta \chi^2$ (blue) since the algorithm tries to interpolate in the mass range between our mass "bins".}. 
However, what is shown in Fig.~\ref{fig:heatmaps} represents only the results 
coming from one background for each of the four channels. We find that both the 95$\%$ upper limits and the 
excess contours depend on the exact astrophysical background used. 
That statement is true for all channels studied and for a wide range of masses. 

We discuss first the impact that alternative astrophysical background 
assumptions have on the upper limits to the dark matter cross section. 
By exploring the parameter space we have found that models with similar 
assumptions 
for the combination of energy losses and diffusion end up giving 
similar upper limits and excess regions in each annihilation channel and type of 
background (low-energy extrapolated or not). That 
is especially true for the energy 
losses in agreement with \cite{Bergstrom:2013jra}. Changing the assumptions on the ISM energy 
loss rate, does not dramatically change the shape of the dark matter
$e^{\pm}$ spectra in the energy ranges that we fit \footnote{For masses 
$m_{\chi} \sim 50$ GeV and for propagated $e^{\pm}$ of energy up to 5 GeV, the 
spectral shapes of the dark matter fluxes studied here change very little by varying the ISM conditions. 
At energies lower than $\sim 5$ GeV, lower energy losses make the dark matter 
spectrum at Earth harder. Given that we focus on relatively light dark matter, 
alternative choices on the ISM energy losses have little impact in the spectral shape (excluding the normalization) of the dark matter originated fluxes.}. 
However, it does affect the amplitude of these fluxes for a given mass, cross 
section and channel. Lower energy losses allow for the observed $e^{\pm}$ of 
dark matter origin to be sourced from a wider volume of the local Milky Way. 
Thus, the dark matter $e^{\pm}$ flux that reaches us is higher when we assume 
lower energy losses; setting in turn tighter limits on the allowed dark matter 
annihilation cross section. Lower ISM energy loss rates enhance in amplitude the 
background astrophysical flux components as well. Yet, that enhancement in those 
fluxes' amplitude is absorbed in our analysis as part of the uncertainty in the 
efficiency of the underlying background sources, i.e the assumptions on the 
SNRs, pulsars and the ISM gas density. Thus changing the energy-loss rate 
assumptions has a residual effect only on the annihilation cross section limits.
Instead, the assumptions about how fast cosmic rays diffuse have a much smaller 
impact on the final dark matter limits. Similarly, the 
assumptions of the galactic scale-height have a very small effect on the 
cross section upper limits. For a given dark matter model's propagated $e^{\pm}$
spectrum, the lower the observed energy is, the larger the volume of origin 
of those $e^{\pm}$ within the Milky Way. For the dark matter masses we focus 
here, different choices on the diffusion coefficient, diffusion index and 
scale height affect mostly the propagated spectra at energies below $\sim 5$ 
GeV, that we do not fit due to the large solar modulation modeling uncertainties. 
The effects of different diffusion assumptions on 
the background fluxes are important if one fixes all other modeling assumptions; 
but once marginalizing over the rest of those assumptions, the diffusion has 
a small 
effect on how much room there is for an additional dark matter $e^{\pm}$ flux 
component. 

\begin{figure*}
\begin{centering}
\hspace{-0.1cm}
\includegraphics[width=3.50in,angle=0]{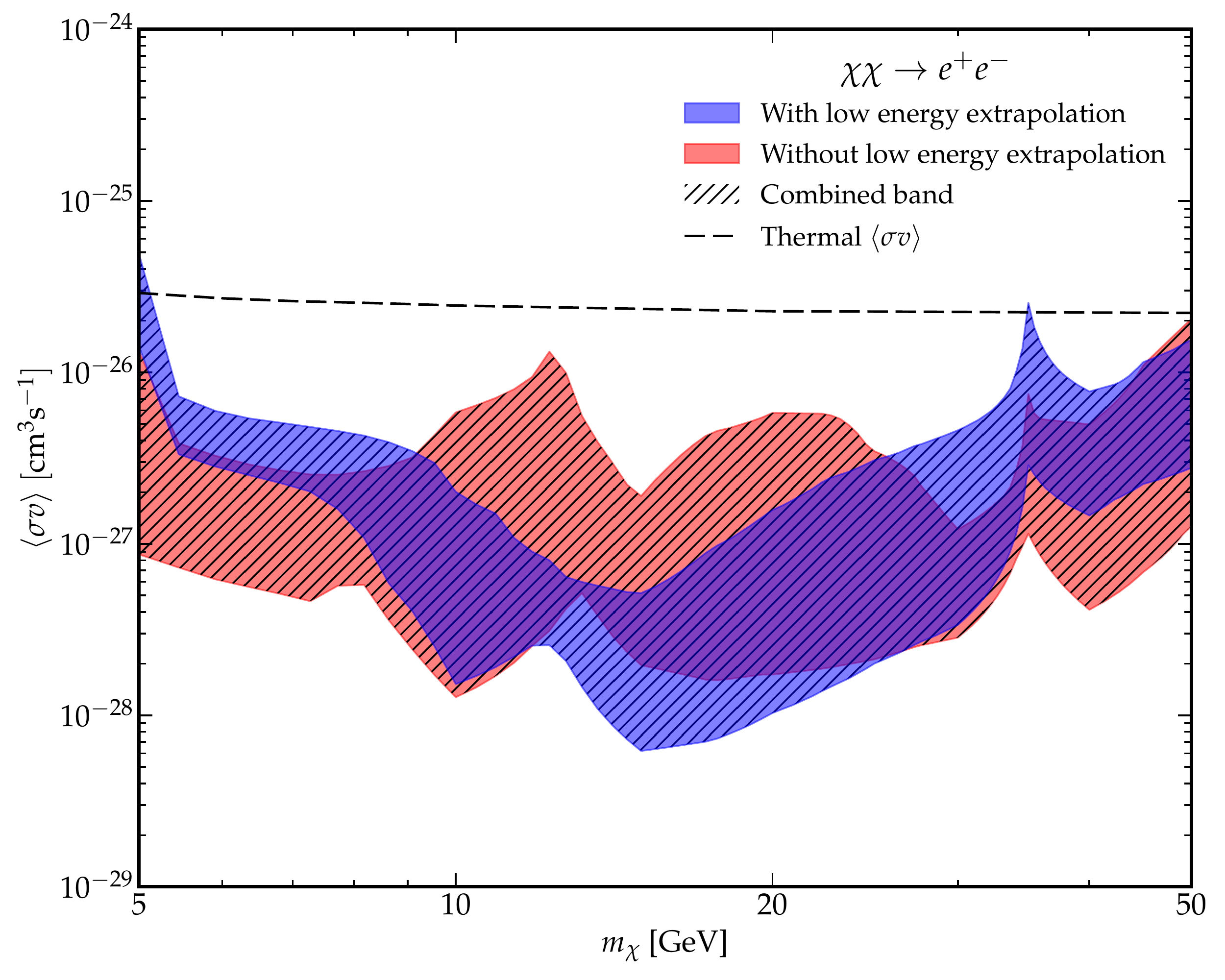}
\includegraphics[width=3.50in,angle=0]{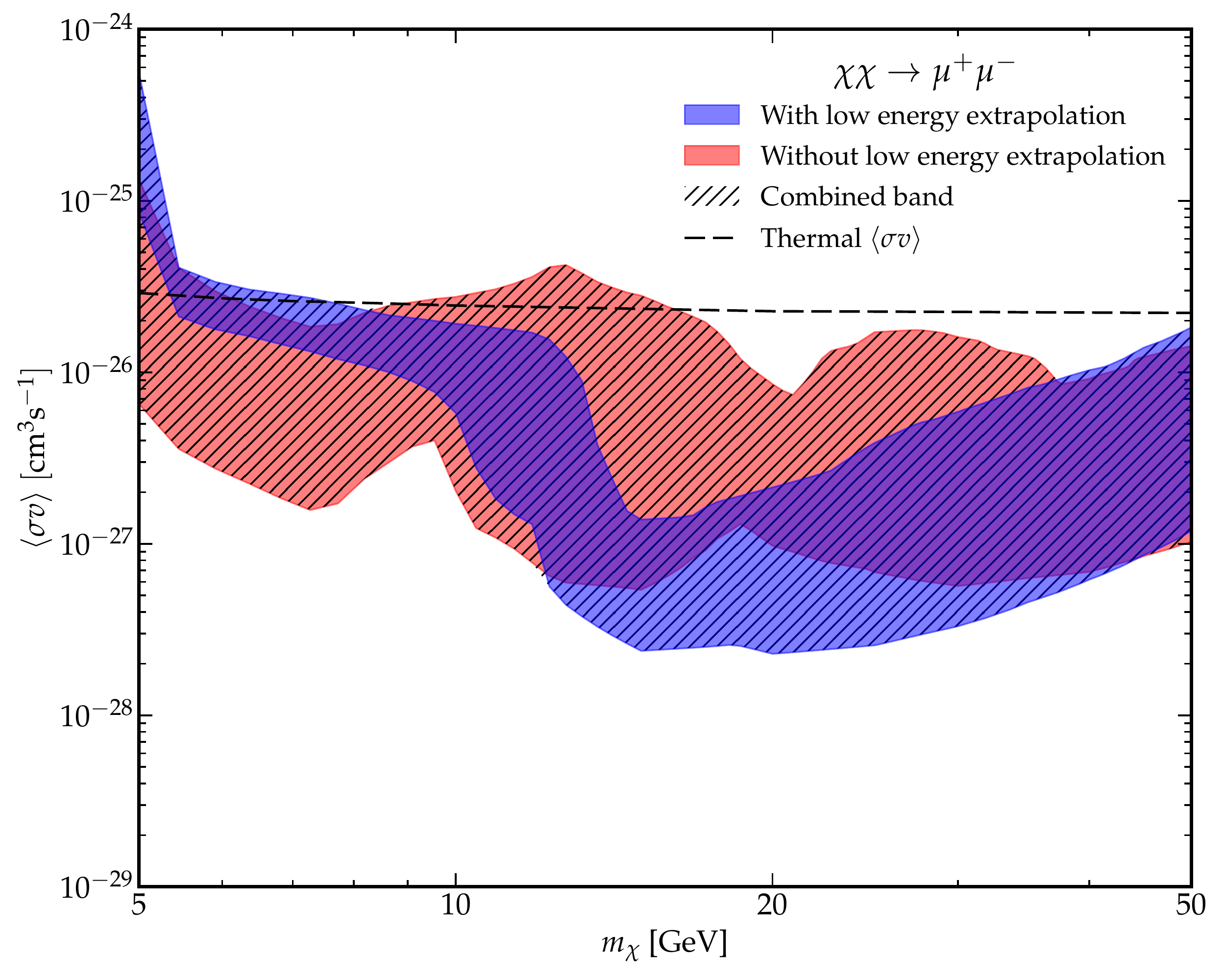}
\end{centering}
\vspace{-0.5cm}
\vspace{-0.0cm}

\begin{centering}
\hspace{-0.1cm}
\includegraphics[width=3.50in,angle=0]{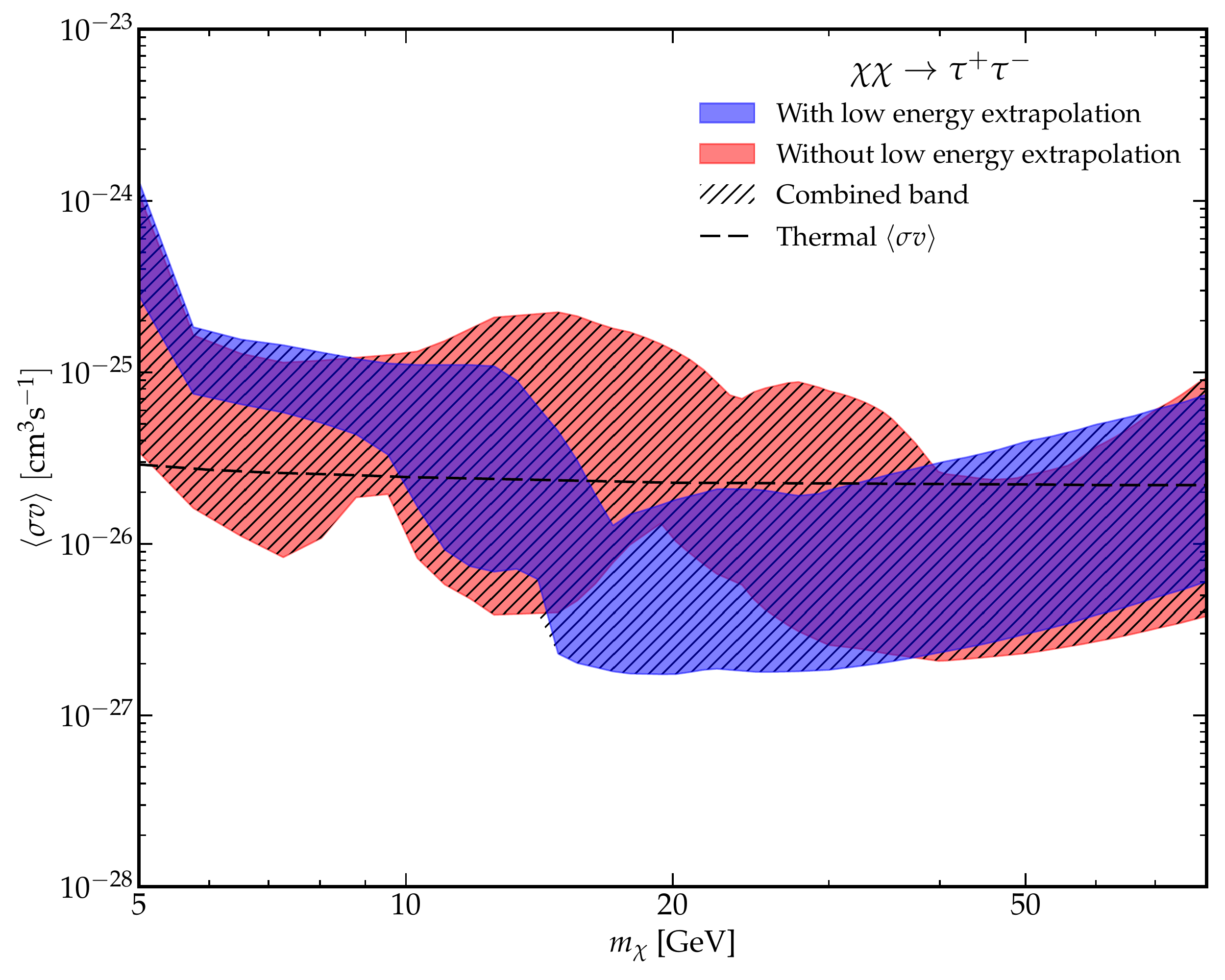}
\includegraphics[width=3.50in,angle=0]{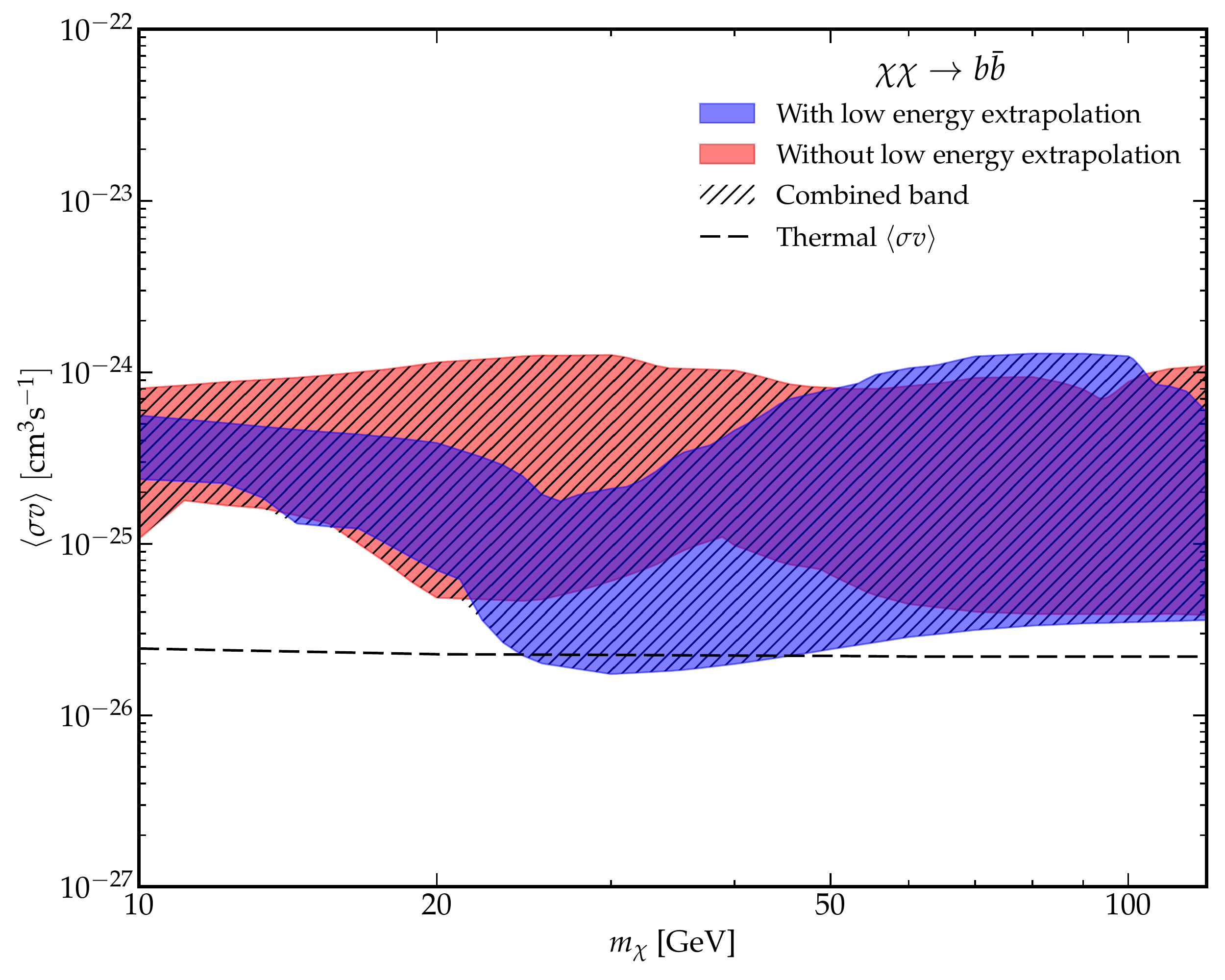}
\end{centering}
\vspace{-0.5cm}
\caption{The 95\% upper limit bands on the annihilation cross section $\langle \sigma v \rangle$ to $e^{+} + e^{-}$ (upper left), $\mu^+ \mu^-$ (upper right), $\tau^+ \tau^-$ (lower left) and $b \bar{b}$ (lower right), each from 60 astrophysical backgrounds that sufficiently cover the astrophysical backgrounds
parameter space.
We separate the bands that come from backgrounds with (blue) and without (red) low energy extrapolation and also show the combined band (hatched).
The dashed lines are the expected $\langle \sigma v \rangle$ for a thermal relic taken from Ref.~\cite{Steigman:2012nb}. 
}
\vspace{-0.0cm}
\label{fig:bands}
\end{figure*}

Having created heatmaps as those of Fig.~\ref{fig:heatmaps}, testing the variety 
of alternative astrophysical assumptions we concluded that instead of using the entire 
library of 1020 astrophysical background assumptions we can reduce our analysis 
to a sample of 60 astrophysical background models for each annihilation channel (there is some overlap between channels).
All of these 60 backgrounds are within $2\sigma$ from a $\chi^2/n_{\mathrm{dof}}=1$ after including dark matter at their best fit point.
Combining different upper limit lines from different backgrounds for the same channel, we construct upper limit bands.
In Fig.~\ref{fig:bands}, we show these upper limit bands for our four annihilation 
channels evaluated from our 60 astrophysical background models per dark matter 
channel.

The 60 background models include the probed range of local energy losses, the 
alternative choices for the Milky Way's local diffusion properties, as well as 
alternative assumptions for the pulsars' flux component at energies of $O(10)$ GeV. 
The energy-loss assumptions affect mostly the $e^{\pm}$ fluxes of the dark matter
component at energies close to the dark matter mass $m_{\chi}$. As the leptonic 
channels give dark matter fluxes to $e^{\pm}$ that are harder in spectrum than the 
primary $e^{-}$ and secondary $e^{\pm}$ astrophysical components,  the dark matter 
component becomes most important in relevant terms at energies close to $\sim 
m_{\chi}$. 
As we explained earlier, alternative choices for energy losses (and to a smaller 
extend diffusion) affect that energy range 
where for a given mass and annihilation channel the dark matter flux becomes most 
relevant. The alternative choices on the pulsars' flux component at energies 
of $O(10)$ GeV are relevant here for the dark matter masses that we study. 
At $O(10)$ GeV many old and distant pulsars contribute and as a result their 
contribution depends on a larger number of assumptions than the contribution of 
local and younger pulsars at higher $e^{\pm}$ energies. We represent that by
breaking our results in Fig.~\ref{fig:bands}, into models ``with low-energy 
extrapolation'' and models ``without low-energy extrapolation''.
The low-energy extrapolation backgrounds give a higher flux from distant and 
older pulsars. The cosmic-ray measurements from \textit{AMS-02}, \textit{CALET} 
and \textit{DAMPE} probe best the pulsars' contribution at high energies. 
Thus, using a range of assumptions that result in a wider range of predictions on the more 
uncertain low-energy astrophysical $e^{\pm}$ background, is the conservative way 
to set limits on annihilating dark matter particles with mass of 5-50 GeV.
In the 60 backgrounds we make sure to include 30 of the best-fit models from 
Ref~\cite{Cholis:2021kqk}, where the full parameter space of local Milky Way 
properties was probed and models for which after adding the dark matter component 
we get high quality fits to the data.

In Fig.~\ref{fig:bands}, there is a rough order in the values of the 
annihilation cross section limits. Our fits show that there is preference for 
a higher $\langle \sigma v \rangle$ when we use $b \bar{b}$ as our channel, followed by 
$\tau^+ \tau^-$, followed by $\mu^+ \mu^-$ and then $e^+ e^-$. 
That is to be expected as the annihilation channels to more massive Standard Model 
particles give $e^{\pm}$ fluxes that span a wider range and have less prominent 
spectral features. This makes the dark matter $e^{\pm}$ fluxes easier to conceal 
within the background fluxes and their modeling uncertainty.

Another result, is that for the three leptonic channels the backgrounds with 
the low-energy extrapolated pulsars' 
flux prefer smaller dark matter masses of the order of 5-10 GeV, while the 
backgrounds without the low-energy extrapolation generally prefer higher masses. 
Following, we explain first the lower masses results. 
Backgrounds with low-energy extrapolation have a higher flux originating 
from pulsars at energies $\lsim O(10)$ GeV 
\footnote{We write at $E \lsim O(10)$ GeV, instead of $E \lsim 10$ GeV, as the low 
energy extrapolation that we use for the modeled pulsar $e^{\pm}$ flux starts at 
energies anywhere between 10 and 30 GeV depending on modeling assumptions.}. 
In our multi-dimensional parameter space minimization, when pulsars predict a 
higher flux at low energies, the positron fraction is fitted better by a higher 
flux from dark matter as well. That counter intuitive statement is due to the
large impact the first few positron fraction data points (at 5-10 GeV) have on 
the fit. 
We remind that the positron fraction error bars at those energies are the smallest (see Fig.~\ref{fig:pf}).
At those low energies, the secondary positrons provide a prominent flux component. 
The secondary cosmic-ray modeling assumptions together with those on solar modulation dominate the fit
between 5 and 10 GeV. 
Changing the pulsars' flux assumptions at energies bellow $O(10)$ GeV 
impacts the secondary positrons and electrons spectra in the entire energy range
of the observed positron fraction.   
The impact of this on the dark matter limits is shown in Fig.~\ref{fig:bands}.
Backgrounds with a 
higher pulsar $e^{\pm}$ flux prediction at low energies result also in more 
room for a \textit{localized in energy} dark matter signal.
Again the "with low-energy extrapolation" backgrounds, predict a higher $e^{\pm}$ 
flux from pulsars at energies $\lsim$ O(10) GeV to the "without low-energy 
extrapolation". For the leptonic channels, that give $e^{\pm}$ flux that peaks 
in a small energy range as is especially the case with the
$\chi\chi \rightarrow e^{+} e^{-}$ and $\chi\chi \rightarrow \mu^{+} \mu^{-}$ 
channels the low energy extrapolation backgrounds usually give weaker limits. 
We note that for the $\chi\chi \rightarrow b \bar{b}$ channel, the situation 
is different as the dark matter flux spans a wide energy range.

The cosmic-ray positron fraction has three spectral features around 12, 21 
and 48 GeV that were identified to be prominent in Refs.~\cite{Cholis:2021kqk} 
and~\cite{Cholis:2022ajr}. For the $\chi\chi \rightarrow e^+ e^-$ channel, the 
lower two in energy features have an effect on our limits. This annihilation 
channel gives a dark matter $e^{\pm}$ flux with a sharp break at energy 
$E \lsim m_{\chi}$.
For the backgrounds without the low-energy extrapolation we see three mass 
regions at $m_{\chi} = 10- 12$ GeV, at $m_{\chi}\sim 20$ GeV and at 
$m_{\chi}\sim 35\ \rm GeV$ where the limits are weaker. Dark matter particles 
with mass of 10-12 GeV, give cosmic-ray $e^{\pm}$ with energies lower than 12 
GeV. Thus, their limits are not directly affected by the 12 GeV positron 
fraction feature. The weakest limits for that mass range are created by 
backgrounds with enhanced local ISM energy losses. 
The weak limits for the mass region of $m_{\chi}\sim 20$ GeV are associated 
to the 12 GeV feature of the positron fraction measurement and are derived from 
ISM backgrounds with conventional local ISM energy losses.
The weaker limits at $m_{\chi} \sim 35$ GeV, are associated to the 21 GeV feature 
on the positron fraction. Both the blue and red shaded regions find weaker 
$\langle \sigma v \rangle$ limits for $m_{\chi} \sim 35$ GeV, as around 20 GeV in $e^{\pm}$ 
cosmic-ray energy their background predictions are similar.

For the annihilation channel to $\mu^+ \mu^-$, the upper limit bands show two 
mass ranges where the limits become weak, at $m_{\chi}$ 10-15 GeV and at
$m_{\chi} \sim 30$ GeV. Again, the first mass range is not affected by the 
presence of the positron fraction spectral features. The weakest limits for 
the 10-15 mass range come from ISM models of increased local energy losses 
and for the "without low-energy extrapolation" background assumption. The 
weaker limits for $m_{\chi} \sim 30$ GeV, originate from the 12 GeV feature. 
The 21 GeV feature has an effect on the highest-end of the shown mass range 

For the annihilation channel to $\tau^+ \tau^-$, the upper limit bands become 
weak, in the wide mass range of $m_{\chi}$ 10-40 GeV. This is similar to what 
was described for the $\chi\chi \rightarrow \mu^+ \mu^-$ channel. 
However, what was two mass ranges for the $\mu^+ \mu^-$ has merged into one. 
The $\tau^+ \tau^-$, has a dark matter $e^{\pm}$ spectrum less localized in 
energy than the $\mu^+ \mu^-$. 

For the $\chi\chi \rightarrow b \bar{b}$ channel, the limits for backgrounds
without the low-energy extrapolation to the pulsars' component are almost always 
weaker for masses $m_{\chi} < 50$ GeV. As we said the $b \bar{b}$ channel gives
a cosmic-ray $e^{\pm}$ flux that spans a wide range. Thus, a suppressed background
positron flux allows for weaker constraints on the dark matter contribution to 
the positron fraction. For the $b \bar{b}$ channel and for mass $m_{\chi}$ of 
50-120 GeV, the limits are roughly the same for the backgrounds with or without 
the low-energy extrapolation. For these higher masses the dark matter originated 
$e^{\pm}$ flux is prominent above the energy range where the extrapolation is 
used.

\begin{figure}
    \centering
    \includegraphics[width=1\linewidth,angle=0]{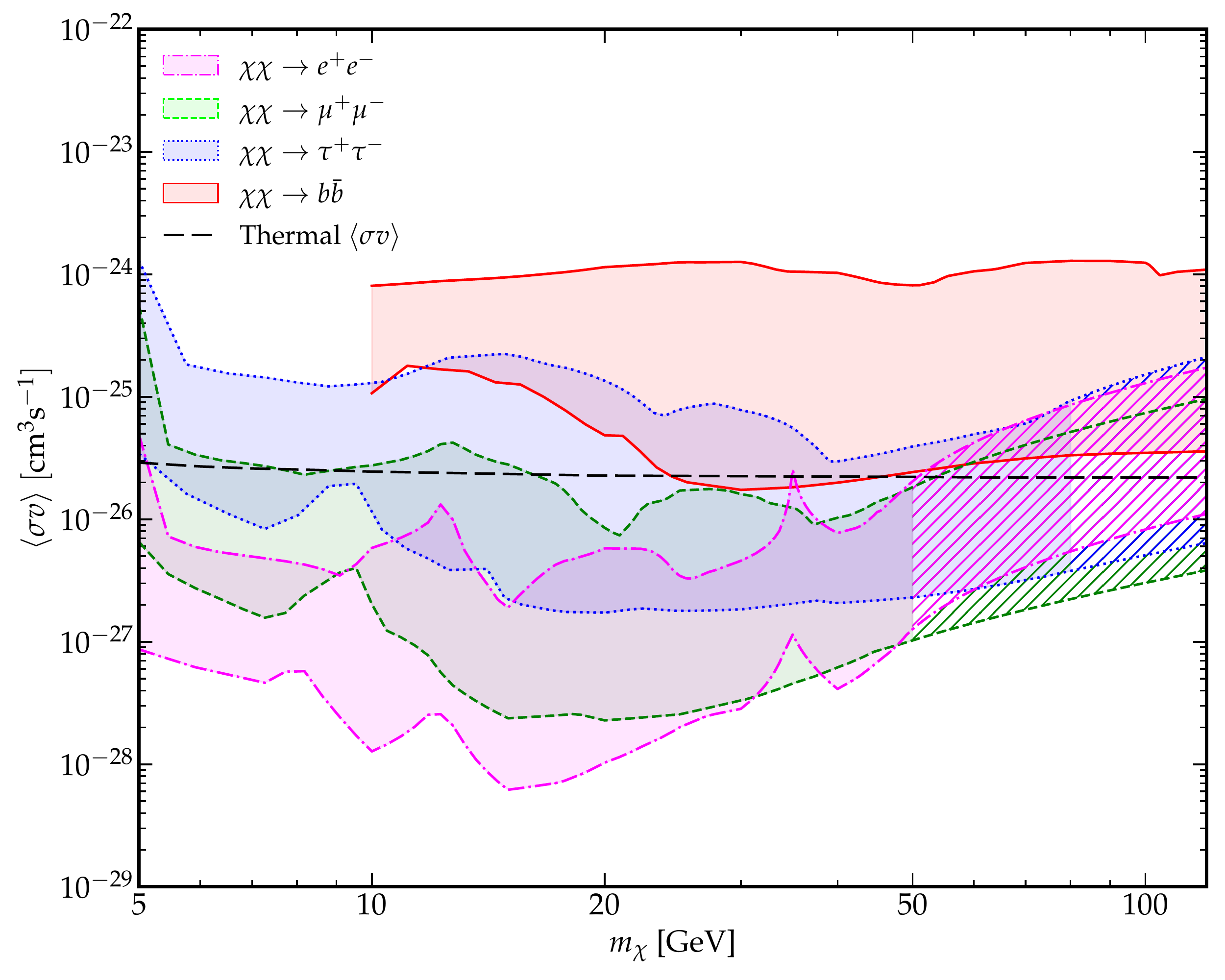}
    \caption{The combined 95\% upper limit bands on $\langle \sigma v \rangle$ for our four different annihilation channels.
    The $e^+ e^-$, $\mu^+ \mu^-$ and $\tau^+ \tau^-$ channel bands are extrapolated up to 120 GeV to match the higher end of the $b \bar{b}$ channel mass range.
    The extrapolated regions are shown hatched.
    The black dashed line is the expected $\langle \sigma v \rangle$ for a thermal relic from Ref.~\cite{Steigman:2012nb}.
    }
    \label{fig:combined_bands}
\end{figure}

In each one of the plots of Fig.~\ref{fig:bands} we also have the combined 
band with and without low energy extrapolation as a hatched region.
In Fig.~\ref{fig:combined_bands}, we show the four combined upper limit bands, one for each of our annihilation channels where we have also extrapolated those bands up to 120 GeV where necessary.
We claim at this point that the upper limits on $\langle \sigma v \rangle$ from 
the cosmic-ray positron fraction, are much more uncertain than previously claimed.
Due to the uncertainties in the astrophysical backgrounds including the pulsar component, there isn't a well defined upper limit line.
Instead, we present upper limit bands evaluated from a collection of viable astrophysical backgrounds, that each gives a different upper limit line.
Our bands at certain masses can span up to two orders of magnitude in the constrained parameter $\langle \sigma v \rangle$, as e.g. for  $\chi\chi \rightarrow \mu^+ \mu^-$ with $m_{\chi} = 15$ GeV.

\subsection{A possible excess flux of low cosmic-ray energy positrons.}
\label{subsec:excess}

\begin{figure*}
\begin{centering}
\hspace{-0.1cm}
\includegraphics[width=3.50in,angle=0]{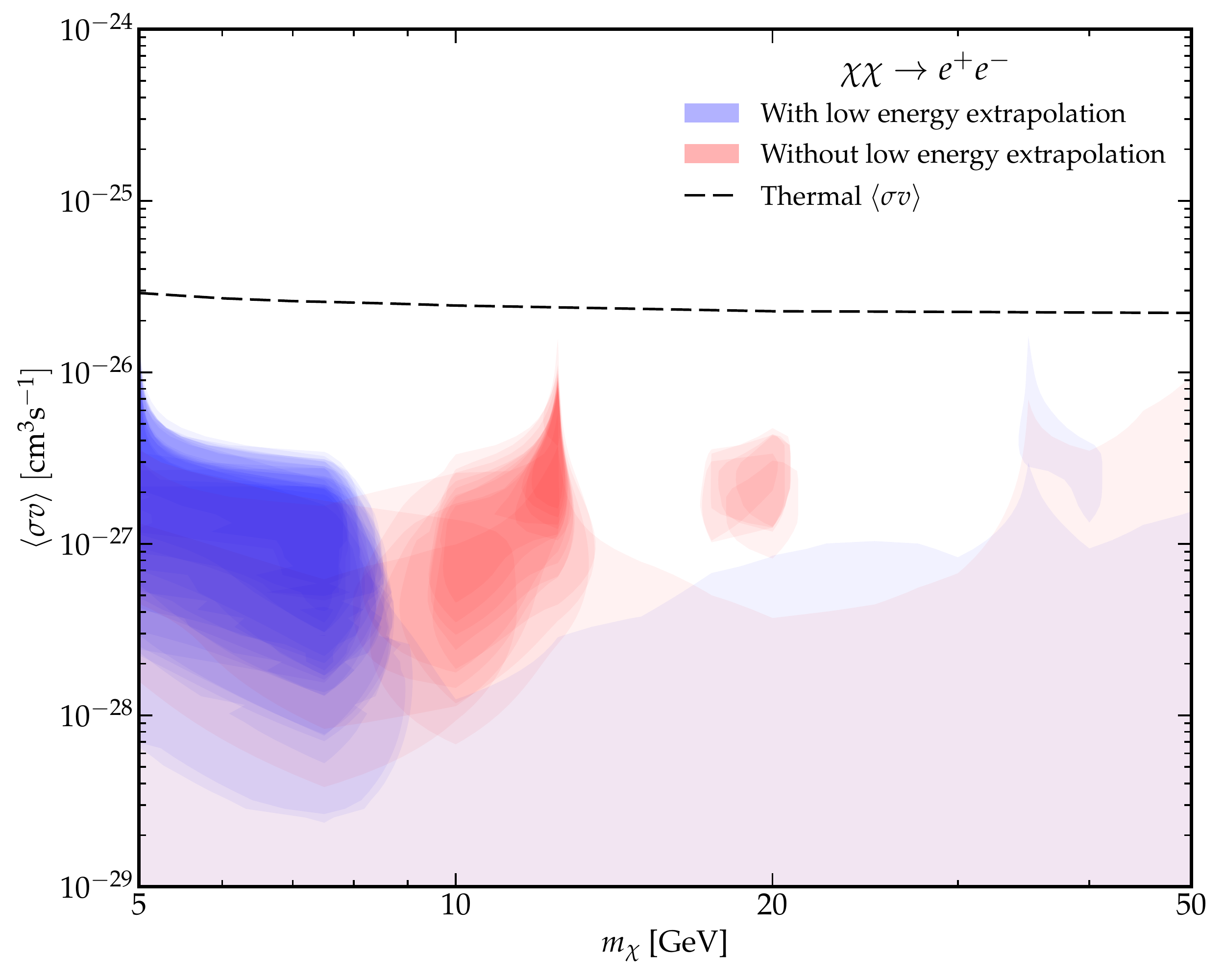}
\includegraphics[width=3.50in,angle=0]{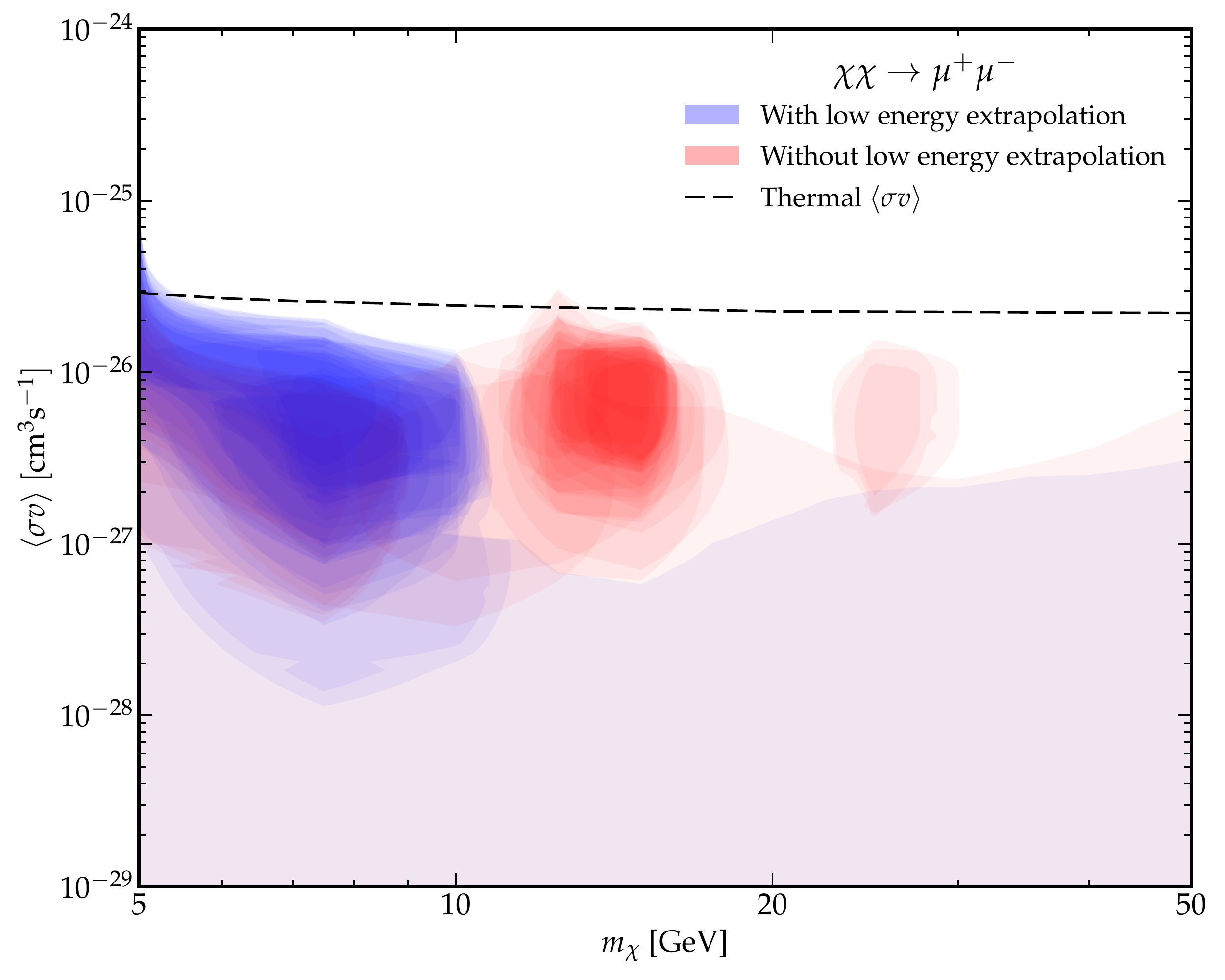}
\end{centering}
\vspace{-0.5cm}
\vspace{-0.0cm}

\begin{centering}
\hspace{-0.1cm}
\includegraphics[width=3.50in,angle=0]{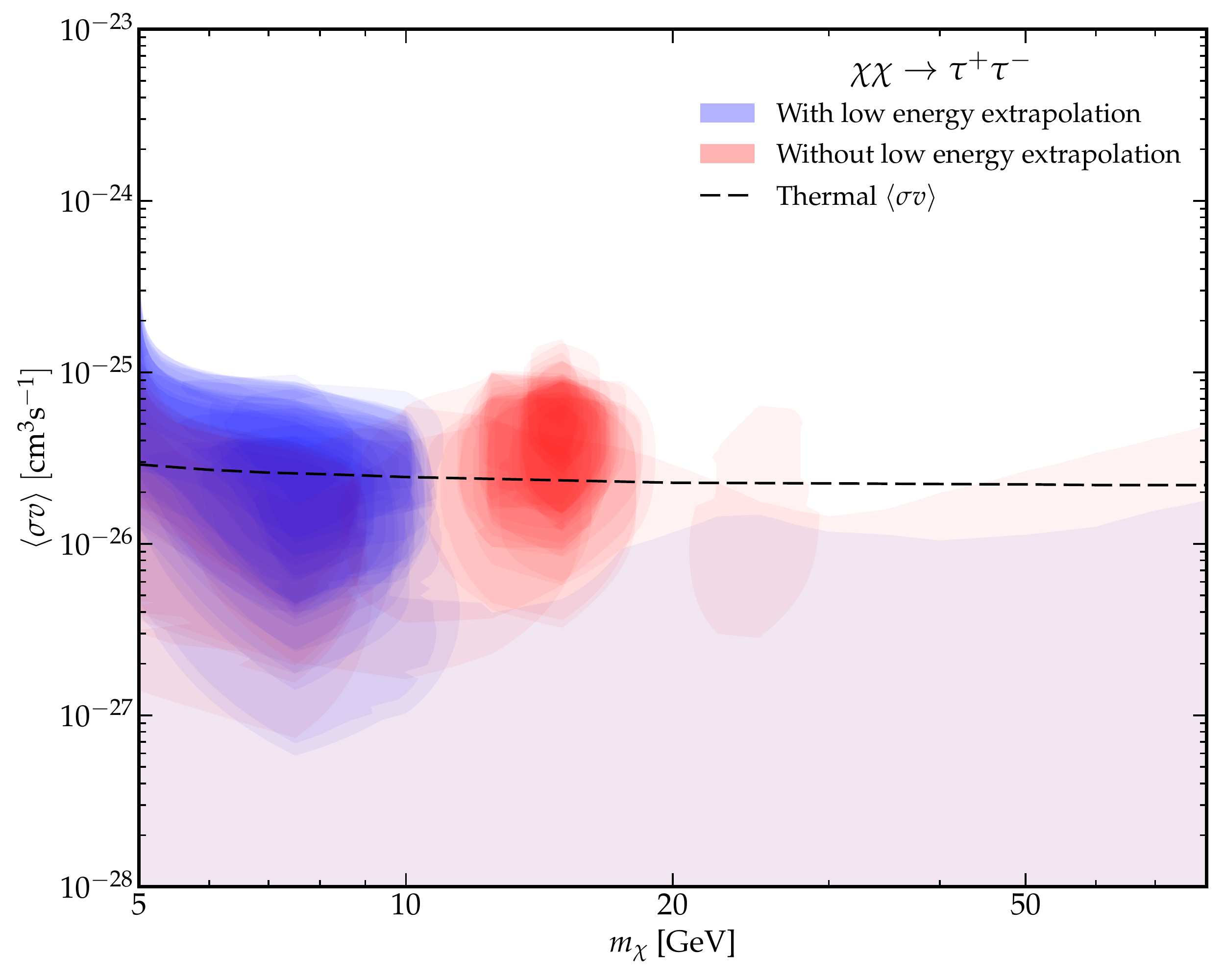}
\includegraphics[width=3.50in,angle=0]{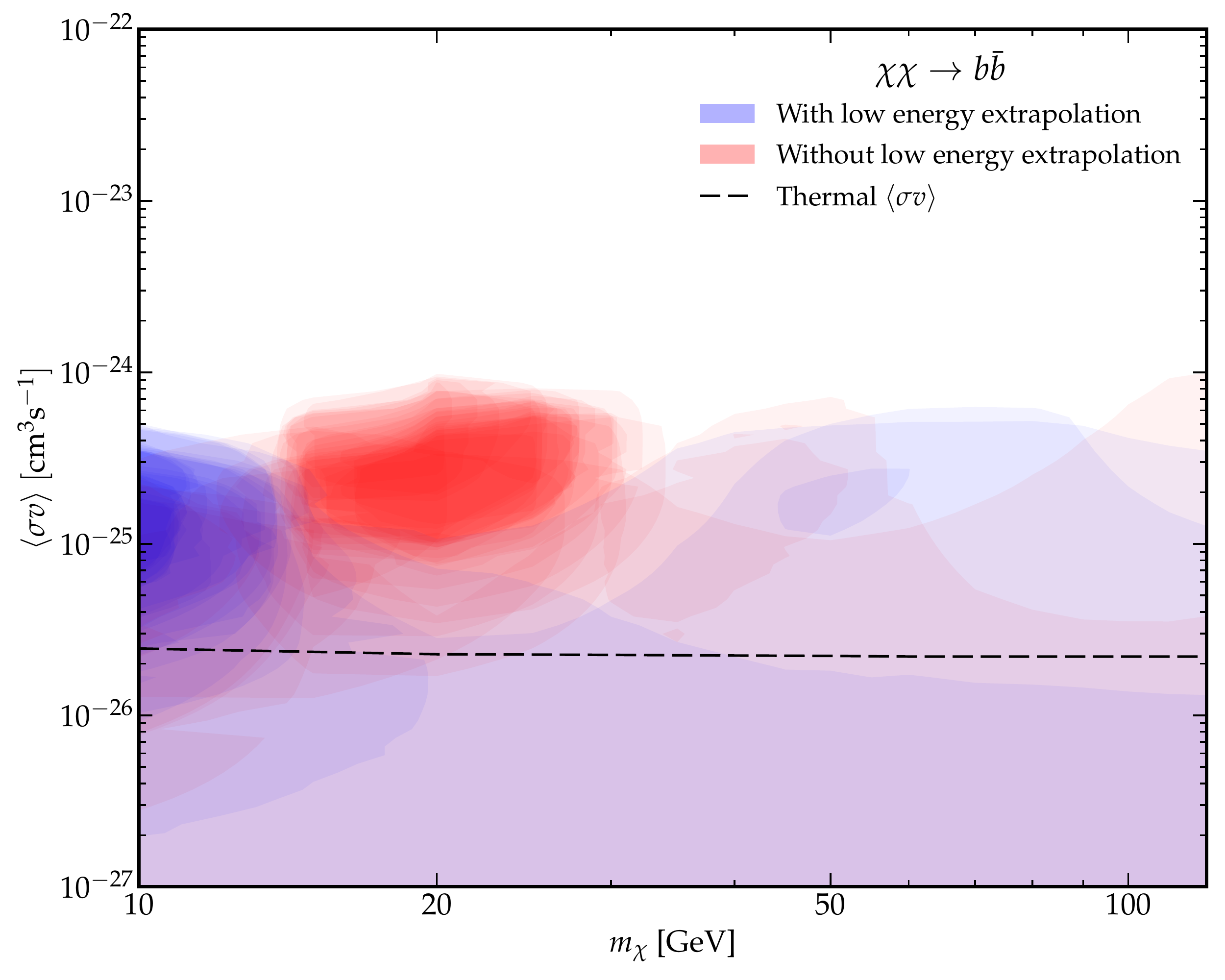}
\end{centering}
\vspace{-0.5cm}
\caption{In a similar manner to Fig.~\ref{fig:bands}, we show the $2\sigma$ excess contour regions for the dark matter annihilation cross section versus mass.
The dashed lines are the expected $\langle \sigma v \rangle$ for a thermal relic taken from Ref.~\cite{Steigman:2012nb}. 
}
\vspace{-0.0cm}
\label{fig:contours}
\end{figure*}

As we show in Fig.~\ref{fig:heatmaps}, for the specific background astrophysical model used, we can find
combinations of dark matter mass, annihilation cross section and channel, where there is 
a significant statistical preference for an additional dark matter component. In that example, the 
fit to the positron fraction improves by a $\Delta \chi^2$ of up to 40. The most statistically 
significant result comes from a $\chi\chi \rightarrow \tau^+ \tau^-$ with $m_{\chi} =7.5$ GeV and
$\langle \sigma v \rangle \simeq 1.5 \times 10^{-26} \ \mathrm{cm^3 s^{-1}}$. Such a dark matter 
particle, has similar properties to those required to explain the galactic center excess in gamma rays 
\cite{Goodenough:2009gk, Vitale:2009hr, Hooper:2010mq, Hooper:2011ti, Gordon:2013vta, Daylan:2014rsa, Calore:2014xka, Calore:2014nla, Abazajian:2014fta, Fermi-LAT:2015sau, DiMauro:2021raz, Cholis:2021rpp}.

In this subsection, we further pursue this indication of a possible dark matter component. We test 
whether this is a result of just a specific combination of astrophysical background assumptions, or 
a result that stands to further scrutiny once we allow for a wide range of astrophysical backgrounds. 
We use the set of 60 astrophysical background models with all the seven astrophysical nuisance 
parameters plus the one related to the dark matter annihilation cross section as we described in 
section~\ref{subsec:analysis}. 

In Fig.~\ref{fig:contours}, we show in a similar manner to Fig.~\ref{fig:bands}, the $2\sigma$ excess 
contour regions for our four dark matter annihilation channels. 
There are 60 overlaid contour regions in each plot.
These excess regions are in agreement 
with our upper limit bands. The respective upper limits like the ones presented in Fig.~\ref{fig:heatmaps} 
for a specific background, get weaker when the excess contours suggest a high annihilation cross section.
Some isolated islands in the cross section versus mass space are also observed. These are created by very 
few background models and are typically of minor statistical significance. In fact, we find that for some 
of the backgrounds used, while there still is some improvement by adding a dark matter component, that is a
small one. In particular, for 2 out of the 60 background models, by adding dark matter annihilating to 
$e^+ e^-$, or $\mu^+ \mu^-$, or $\tau^+ \tau^-$ leptons the improvement in $\Delta \chi^2$ is less than 6.
That results in open $2\sigma$ contours depicted by light red or blue regions extending to effectively zero
annihilation cross sections. 
For dark matter particles annihilating to $b \bar{b}$, 
there are 6 out of the 60 backgrounds where that is the case. This shows that this ``excess'' is not always
statistically significant. However, we note that for most of the tested backgrounds there is some excess
$e^{\pm}$ flux compatible to that coming from locally annihilating dark matter. As we will explain 
in detail in the following subsection~\ref{subsec:mock}, even if indeed this is a dark matter signal,
deducing its particle properties is highly challenging. 
Also, we note that some of the parameter space suggested by the excess is excluded by limits 
from the cosmic microwave background, gamma rays and antiprotons \cite{Planck:2018vyg, Fermi-LAT:2016uux, Cholis:2019ejx}.

We can not claim and do not claim any robust $e^{\pm}$ excess flux between 5 and 15 GeV on the 
positron fraction spectrum. We just note that we used background models that always agree with higher 
energy $e^{\pm}$ observations from \textit{AMS-02}, \textit{CALET} and 
\textit{DAMPE}, and that half of the 60 astrophysical background models give 
(without a dark matter component) a good quality fit of $\chi^2/n_{\mathrm{dof}} < 1.3$ to the positron fraction from 5 GeV to 1 TeV. 
Yet, a $>2 \sigma$ 
excess for the low energy cosmic-ray $e^{\pm}$ flux is found for the great majority of these background models. We also find a tendency for background 
models with high local energy losses, of $dE/dt \sim 8 \times 10^{-6} \, \textrm{GeV}^{-1} \, \textrm{kyr}^{-1} (E/(\textrm{1 GeV}))^{2}$, to reduce the significance for that excess component. 
Further understanding of the astrophysical, nuclear and particle physics uncertainties at the 1-20 GeV 
cosmic-ray energy range is needed to more accurately model the primary and secondary production of cosmic rays, 
as well as constrain better their propagation properties through the ISM and the heliosphere. Finally,
having a full knowledge of the correlated errors on the positron fraction measurement will be of great
benefit to elucidate the \textit{AMS-02} observations. In the absence of such a published result by the
\textit{AMS-02} collaboration, we leave this issue of a possible $e^{\pm}$ ``excess'' flux an open 
question. 

\subsection{Ability to detect dark matter in the \textit{AMS-02} positron fraction measurement}
\label{subsec:mock}

\begin{figure*}
\begin{centering}
\hspace{-0.1cm}
\includegraphics[width=3.50in,angle=0]{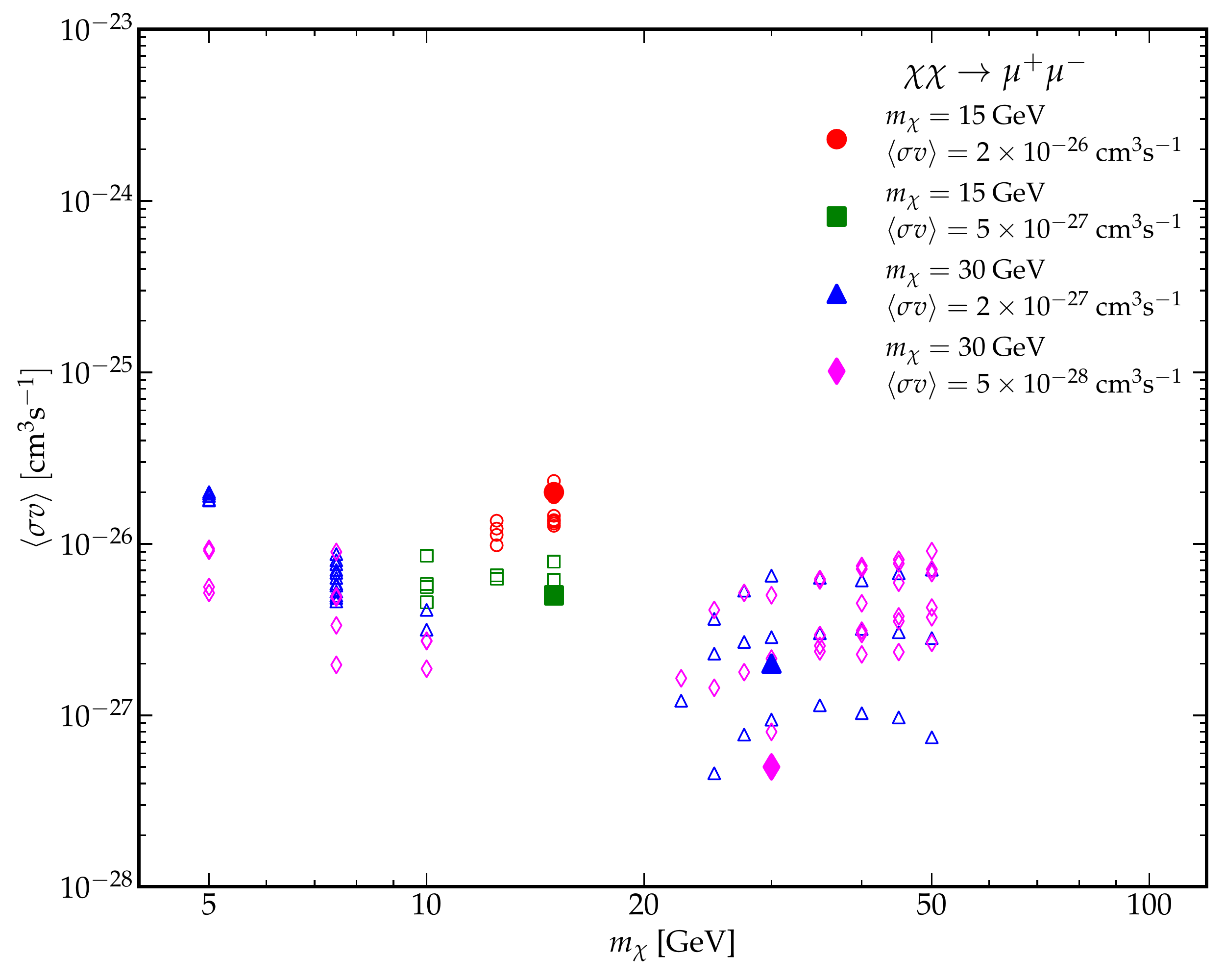}
\includegraphics[width=3.50in,angle=0]{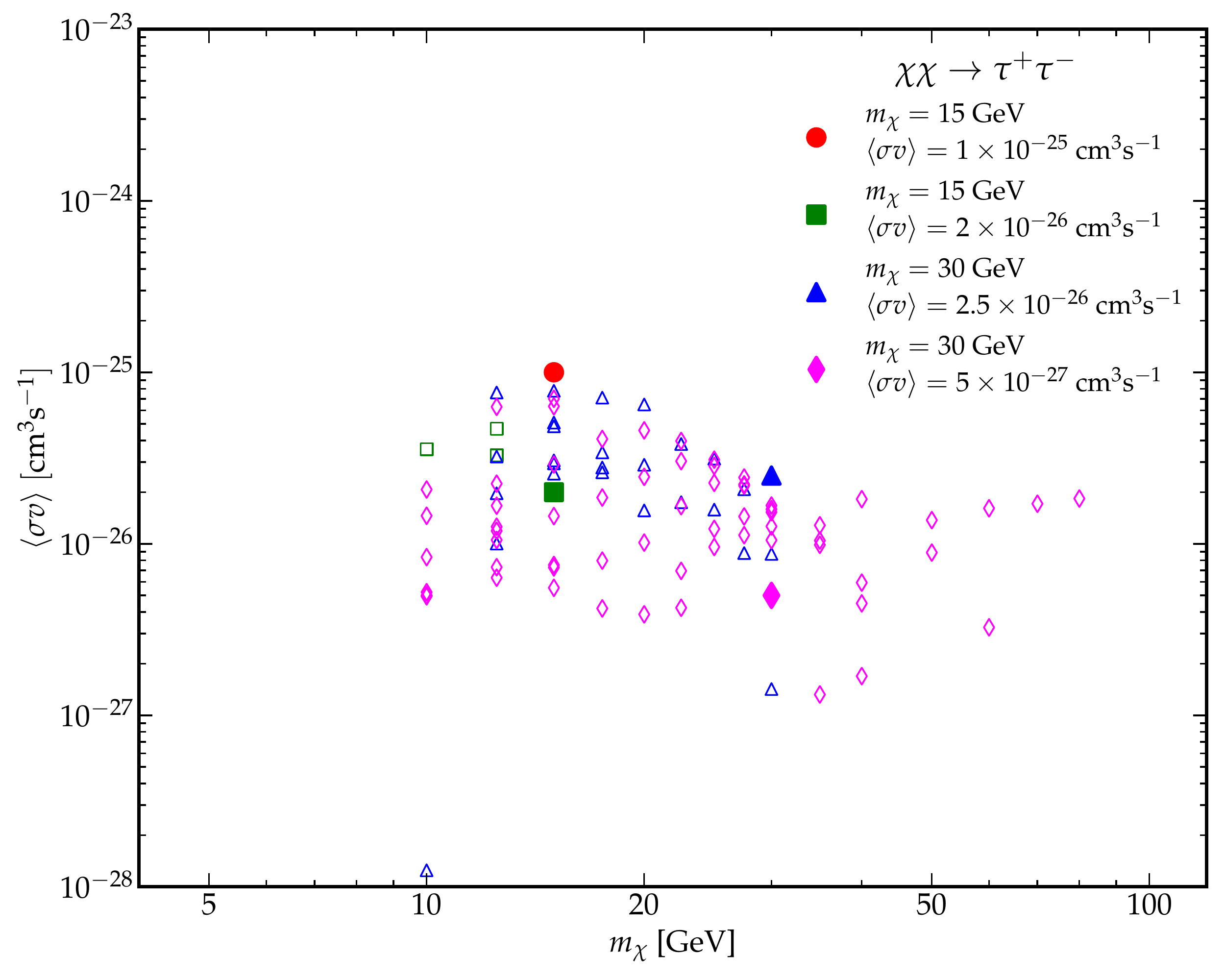}
\end{centering}
\vspace{-0.5cm}
\caption{Scatter plots of best fit points in our 2D dark matter parameter space of the backgrounds + dark matter models that are within $2\sigma$ of a $\chi^2/n_{\mathrm{dof}}=1$. We show results derived after scanning the mock data testing only the same dark matter channel to the one used to create them.
On the left, we show the results for the $\chi \chi \rightarrow \mu^+ \mu^-$ mock data, while on the right, we show the results for the $\chi \chi \rightarrow \tau^+ \tau^-$.
With filled larger markers we denote the original points, i.e. the dark matter properties that we used to create the mock data. With the smaller empty markers of the same style, we denote the best fit points derived from scanning those mock data.}
\vspace{-0.0cm}
\label{fig:mock_data}
\end{figure*}

In this section if the \textit{AMS-02} positron fraction 
measurement can be used to probe the properties of a dark matter component in it. 
As we describe in section~\ref{subsec:mockdata}, we use eight combinations of 
dark matter mass, annihilation cross section and channel to create mock positron fraction measurements. 
These mock positron fraction measurements take also into account the flux 
predictions from one specific background per annihilation channel. 
These two backgrounds were selected to provide a good fit to the \textit{AMS-02}, \textit{CALET} and \textit{DAMPE} cosmic-ray lepton measurements (see Ref.~\cite{Cholis:2021kqk}). 

We test these mock observations by searching for a dark matter signal. 
We first want to test if by adding an excess signal of dark matter origin we 
can at the very least identify its presence, i.e. the presence of an additional 
term that doesn't come from cosmic-ray primaries, secondaries or the main 
population of local pulsars. That doesn't exclude that an excess flux in 
low-energy positrons can still be of a more conventional astrophysical origin as 
e.g. a collection of very powerful, nearly simultaneous but old supernova remnants 
giving a high flux of $e^{\pm}$ in that range \cite{Mertsch:2018bqd}, or a similar 
collection of very (and atypically) powerful pulsars, or $e^{\pm}$ from  cosmic-ray activity in the 
center of the Milky Way \cite{Cholis:2022ajr}. As a second level question, we want 
to see if assuming that excess term is of dark matter origin, whether we can 
identify its specific particle physics properties (mass, annihilation channel and 
cross section). 

We fit each one out of the eight mock data by being agnostic to both the 
background and the dark matter signal. We test all combinations of 100 
backgrounds and all dark matter channels and masses  (64 in total). 
Thus we perform $6400 \times 8$ such fits. We check whether the combinations of 
background $\&$ dark matter channels and masses that were used to create the mock 
data are statistically significant against the other combinations. 
We remind the reader that the annihilation cross section is let to be free in 
our fitting procedure. 
We find that the astrophysical background is easily identifiable i.e. we can find
which one out of the 100 backgrounds was used to create the mock 
positron fraction. 
We also find that we can at the very least identify the presence of an 
additional term contributing to the positron fraction, i.e a nonzero annihilation 
cross section.  
When fitting to mock data that originated from a particular astrophysical 
background, all the good fits found are combinations of that background 
$+$ some dark matter contribution.
However, the properties of that dark matter component are not easily identifiable.
As the astrophysical background dominates the positron measurement in the entire 
energy range it is possible to pick out the underlying background model used to 
create the mock data. That is not the case with the dark matter component that 
affects the measurement only in a narrow energy range.  

The dark matter annihilation channel is very challenging to  
identify. After scanning our mock data, we find that there is no statistical significance for 
models that are built with the original dark matter annihilation channel over the other three channels. 
Among the best fits to each mock data, 
the fit doesn't give any preference between different annihilation channels. 
Even if a dark matter signal is 
present within cosmic-ray $e^{\pm}$, we cannot identify the annihilation channel 
that produces that signal. For that reason, we perform an additional test where we 
fix the dark matter annihilation channel to the one used to create each 
of our mock data.

In Fig.~\ref{fig:mock_data}, we show the results of scanning our mock data 
with only the channels that were used to create them. On the left we show the 
results for the $\chi \chi \rightarrow \mu^+ \mu^-$ mock data and on the right the 
results for the $\chi \chi \rightarrow \tau^+ \tau^-$ mock data.
For each one of these eight cases we test a combination of 100 backgrounds 
$\times$ 17 masses. The maximum number of empty markers that could exist for 
each type is 1700. Those that appear in the plots are the ones within $2\sigma$ 
of a $\chi^2/n_{\mathrm{dof}}=1$, which means $\chi^2/n_{\mathrm{dof}} \leq 1.31$.
If small clusters of similar empty markers (best fit points) form around the 
same filled markers (original mock data parameters), then this points to our 
ability to correctly identify the dark matter properties using the positron 
fraction. We stress that for these results, we are not being agnostic about 
the annihilation channel. To know the annihilation channel further input would 
be required. 

For the $\chi \chi \rightarrow \mu^+ \mu^-$ case (left), the results depend on 
the mass of the dark matter particle. When the mock data contain a signal with 
$m_{\chi} = 15\ \rm GeV$ (red circles and green squares), we can identify the 
properties of this signal relatively well. Our fits give masses and cross 
sections very close to the original ones. However, when the mock data contain 
a signal with $m_{\chi} = 30\ \rm GeV$ (blue triangles and magenta diamonds), the 
signal becomes unidentifiable. Our fits give masses that span between 5 and 50 
GeV which is our entire tested mass range. Also, the cross section is very 
uncertain, spanning more than one order of magnitude around the original value.
For the $m_{\chi} = 30$ GeV case, we used a smaller annihilation cross section 
to create the mock data. The presence of such dark matter doesn't deform the 
positron fraction spectrum significantly and our fitting procedure can adjust 
all the other astrophysical background parameters to fit the mock positron 
fraction very well. That is why for the more massive cases, the dark matter 
signal becomes unidentifiable. For the 
$m_{\chi} = 30$ GeV mock data, we picked the lower annihilation cross sections 
to $\mu^+ \mu^-$, compared to the $m_{\chi} = 15$ GeV, as the relevant limits 
presented in Fig.~\ref{fig:bands}, are stronger for the 30 GeV mass. 

For the $\chi \chi \rightarrow \tau^+ \tau^-$ case (right), it is much harder 
to identify the dark matter properties even when the mass is 15 GeV. 
When $m_{\chi}=15\ \rm GeV$ the cross sections that we decided to use to produce 
the mock data, deforms the positron fraction too much in the lowest 
energies where the error bars are the smallest. Our fits cannot adjust the 
astrophysical background parameters enough to fit this deformation.
The end result is that for $\langle \sigma v \rangle = 1\times 10^{-25}\ \rm cm^3 s^{-1}$ (red 
circles), no fits survive within $2 \sigma$, while for 
$\langle \sigma v \rangle = 2\times 10^{-26}\ \rm cm^3 s^{-1}$ 
(green squares), only 3 fits are within $2 \sigma$. These 3 fits however 
fall around the original mass and the cross section that was used to create the 
mock data so our signal is identifiable within some uncertainty.
When $m_{\chi}=30\ \rm GeV$, the results are similar to the $\chi \chi \rightarrow \mu^+ \mu^-$ 
case and the mass and annihilation cross section of the dark matter particle are 
completely unidentifiable for the same reasons.

\section{Discussion and Conclusions}
\label{sec:conclusions}

In this paper we have used the most recent positron fraction measurement from 
\textit{AMS-02} to set upper limits on dark matter annihilation cross 
sections for masses of 5-120 GeV. Most of the electrons and positrons observed
by \textit{AMS-02} do not originate from dark matter.
Our astrophysical background modeling relies heavily on our 
previous work of Ref.~\cite{Cholis:2021kqk}. The astrophysical backgrounds 
produced in that work are in 
agreement with a wide range of cosmic-ray measurements such as the positron 
fraction, the positron flux and the electron plus positron flux.

Our background models account for cosmic-ray primary electrons and 
uncertainties in their injection amplitude and spectrum, cosmic-ray secondary
electrons and positrons and equivalent uncertainties in their injection 
amplitude and spectrum. They also include the contribution of Milky Way pulsars
that come with a sequence of uncertainties. Those uncertainties are related 
to the stochastic nature
of their (neutron star) birth, distribution in space, initial spin-down power and 
the injected into the ISM cosmic-ray electron/positron fluxes. Pulsars also have
a relatively uncertain time evolution as they spin down and also poorly determined 
efficiency in converting their rotational energy into cosmic-ray electrons and 
positrons. All these uncertainties are accounted for in our background 
simulations. Finally, we test a sequence of alternative modeling assumptions 
on how cosmic-ray electrons and positrons propagate through the ISM and the 
heliosphere, before reaching the \textit{AMS-02} detector. In these background 
models we added a contribution from annihilating dark matter.

We study four different annihilation channels. These are $\chi\chi \rightarrow e^+ e^-$, 
$\chi\chi \rightarrow \mu^+ \mu^-$, $\chi\chi \rightarrow \tau^+ \tau^-$ and 
$\chi\chi \rightarrow b\bar{b}$. We use a discretized grid of dark matter masses $m_{\chi}$ and test them over our large sample of astrophysical backgrounds. In our analysis the annihilation 
cross section, in units of $\langle \sigma v \rangle$ is treated as a free parameter.
We derive upper limits on $\langle \sigma v \rangle$ as a function of $m_{\chi}$ for each 
annihilation channel which are summarized in Fig.~\ref{fig:combined_bands}.

We find that that each valid astrophysical background that explains the \textit{AMS-02} data well, gives different upper limits on $\langle \sigma v \rangle$.
Therefore, we claim that the correct upper limits are not lines but bands spanning roughly one order of magnitude for each annihilation channel.
We also find that the limits become stronger or weaker depending on the dark matter particle mass, and that this, in the $\chi\chi \rightarrow e^+ e^-$ channel, has to do with low energy spectral features in the positron fraction. 
To a weaker extend a similar trait exists for the $\chi\chi \rightarrow \mu^+ \mu^-$ and $\chi\chi \rightarrow \tau^+ \tau^-$ channels. Instead, the limits 
derived for the $\chi\chi \rightarrow b \bar{b}$ case are fairly insensitive 
to the existence of low energy features in the positron fraction spectrum. 
We believe that a better understanding of the secondary production of cosmic 
rays in our Galaxy and solar modulation will help us to further refine these 
dark matter upper limits and reduce the widths of the presented bands.

Furthermore, we note that in the great majority of the astrophysical 
backgrounds that we use which are compatible with the $e^{\pm}$
observations by \textit{AMS-02}, \textit{CALET} and \textit{DAMPE}, we
still find preference for an additional cosmic-ray component. 
That component of 5-15 GeV $e^{\pm}$ is compatible with a dark matter
contribution. However, its statistical significance varies between 
backgrounds and in a small fraction among them the preference for 
dark matter becomes negligible. We do not think that ``excess''
to be robust due to astrophysical background modeling uncertainties.
The \textit{AMS-02} error bars at the the low-end of the energy range that we use are of the order of ~0.5 \%.
That is why the presence of a dark matter flux that contributes at the few percent level to the positron fraction can have a significant impact while fitting to the data.
With future cosmic-ray measurements as for instance \textit{AMS-02} or an 
even more sensitive future cosmic-ray detector \cite{Schael:2019lvx}, we will be able to reduce these
uncertainties and further scrutinize the positron fraction at that
energy range.

Additionally, we created mock data of the \textit{AMS-02} positron fraction to
check whether we could detect a dark matter signal contributing to its spectrum 
and also identify that signals' particle physics properties.
We find that if dark matter contributes approximately 2-3 $\%$ of the positron fraction in at 
least a few low-energy bins, we would be able to recognize the presence of some 
additional term in the \textit{AMS-02} data. However, determining the annihilation 
channel of such a signal is very challenging.
If in such an analysis we allow for knowledge of the 
underlying annihilation channel, then for certain masses it is possible to accurately identify the dark matter properties of our injected mock dark matter signal.

We find that for dark matter annihilating to $\mu^+ \mu^-$, for masses 
$m_{\chi} \sim 15$ GeV and cross sections $\langle \sigma v \rangle \sim 10^{-26} \rm cm^3 s^{-1}$, we can identify the properties of mock signals within some statistical uncertainty.
For larger dark matter masses, the properties of the mock signals to $\mu^+ \mu^-$ 
become more challenging.
For dark matter annihilating to $\tau^+ \tau^-$, we find that for masses $m_{\chi} \sim 15$ GeV and cross sections of the order of $\sim 1\times 10^{-25}\ \rm cm^3 s^{-1}$, it is difficult to fit to the mock data with our models. 
For larger masses, the situation is similar to the $\mu^+ \mu^-$ case where the properties of the dark matter signal are very uncertain.

We have made publicly available our dark matter fluxes in their pre-fitted format for the entire set. These files can be found at \url{https://zenodo.org/record/7178634}. Our astrophysical background models of Ref.~\cite{Cholis:2021kqk}, are instead available at \url{https://zenodo.org/record/5659004}.

\acknowledgements
We thank Dan Hooper for useful discussions and Tim Linden and Isabelle John for comments on the manuscript.
We also acknowledge the use of \texttt{GALPROP} \cite{galpropwebsite, galprop} and
the \path{Python} \cite{10.5555/1593511} modules, \path{numpy} \cite{harris2020array},
\path{SciPy} \cite{2020SciPy-NMeth}, \path{pandas} \cite{reback2020pandas,mckinney-proc-scipy-2010}, \path{matplotlib} \cite{Hunter:2007}, \path{Jupyter} \cite{Kluyver2016jupyter}, and \path{iminuit} \cite{iminuit,James:1975dr}.
IC acknowledges support from the Michigan Space Grant Consortium, NASA Grant No. 80NSSC20M0124.
IC acknowledges that this material is based upon work supported by the U.S. Department of Energy, Office of Science, 
Office of High Energy Physics, under Award No. DE-SC0022352.

\appendix

\section{Summary of the full parameter space}
\label{app:table}

\begin{table*}
    \centering
    \begin{tabular}{|c|c|}
        \hline
         Parameter & Range \\
         \hline \hline
         Galactic halo height, $z_L\ [\mathrm{kpc}]$ &  3.0, 5.5, 5.7, 6.0\\
         Energy loss rate, $b\ [\mathrm{10^{-6}GeV^{-1}kyrs^{-1}}]$ & 2.97, 5.05, 8.02\\
         Diffusion coefficient, $D_0\ \mathrm{[pc^2 kyr^{-1}]}$ & 33.7, 51.3, 92.1, 140.2\\
         Diffusion spectral index, $\delta$ & 0.33, 0.40, 0.43, 0.50\\
         Pulsar braking index, $\kappa$ & 2.5, 2.75, 3.0, 3.25, 3.5\\
         Pulsar characteristic spin-down timescale, $\tau_0\ [\mathrm{kyr}]$ & 0.6, \ldots, 33\\
         Pulsar initial spin-down power cutoff, $x_{\mathrm{cutoff}}$ & 38.0, \ldots, 39.3\\
         Pulsar initial spin-down power mean, $\mu_y$ & 0.1, \ldots, 0.6\\
         Pulsar initial spin-down power standard deviation, $\sigma_y$ & 0.25, \ldots, 0.75\\
         Pulsar cosmic-ray $e^{\pm}$ injection spectral index, $n$ & uniform distribution $\in$ [1.4,1.9], [1.6,1.7], [1.3,1.5]\\
         Pulsar mean efficiency $\bar{\eta}$ and $\zeta = 10^{\sqrt{\mathrm{variance}}}$ of $\bar{\eta}$ in $e^{\pm}$ pairs & $(4\times 10^{-3},1.47)$, $(1\times 10^{-3},2.85)$, $(1\times 10^{-2},1.29)$\\
         Prim. $e^-$ flux normalization, $a$ from reference value (see Ref.~\cite{Cholis:2021kqk}) & $0.6$ - $1.2$\\
         Sec. $e^{\pm}$ flux normalization, $b$ from reference value (see Ref.~\cite{Cholis:2021kqk}) & $0.8$ - $2.0$\\
         Pulsar $e^{\pm}$ flux normalization, $c$ & such that $\bar{\eta} \cdot c \leq 0.5$ \\
         Prim. $e^-$ flux spectral index modifier, $d_1$ from reference value (Ref.~\cite{Cholis:2021kqk}) & $-0.2$ - $0.5$\\
         Sec. $e^{\pm}$ flux spectral index modifier, $d_2$ from reference value (Ref.~\cite{Cholis:2021kqk})& $-0.1$ - $0.1$\\
         Solar modulation parameter, $\phi_0\ [\mathrm{GV}]$ & $0.1$ - $0.6$\\
         Solar modulation parameter, $\phi_1\ [\mathrm{GV}]$ & $0.0$ - $2.0$\\
         Dark matter $e^{\pm}$ flux normalization, $f$ & $0.0$ - $\infty$\\
         \hline
    \end{tabular}
    \caption{List of parameters in our models. The last eight parameters are free fitting parameters and are optimized in our minimization procedure. The rest of the parameters are fixed and are different for each astrophysical pulsar background that we use. For more details on these parameters, refer to sections II and III of Ref.~\cite{Cholis:2021kqk}.
    The last eight parameters are ranges of values while the rest are discreet sets separated by commas.}
    \label{tab:params}
\end{table*}

In this appendix we present briefly the parameter space that we test with our astrophysical background models and provide the ranges that the nuisance parameters are allowed to take.  
This parameter space can be seen in Table~\ref{tab:params}.
The last eight parameters in this Table are free fitting parameters that are optimized while fitting our simulations while the rest of the parameters are fixed and each astrophysical background model has a unique set of pulsar parameters. Not all combinations of astrophysical parameters are allowed based on various observations from pulsars or cosmic rays. For more details on the properties of the constructed astrophysical backgrounds 
and the parameters describing them, please refer to sections II and III of Ref.~\cite{Cholis:2021kqk}.
The last eight parameters are presented as ranges while the rest are discreet sets separated by commas.

\section{Different statistical procedure}
\label{app:diff_stats}

\begin{figure*}
\begin{centering}
    \hspace{-0.1cm}
\includegraphics[width=3.50in,angle=0]{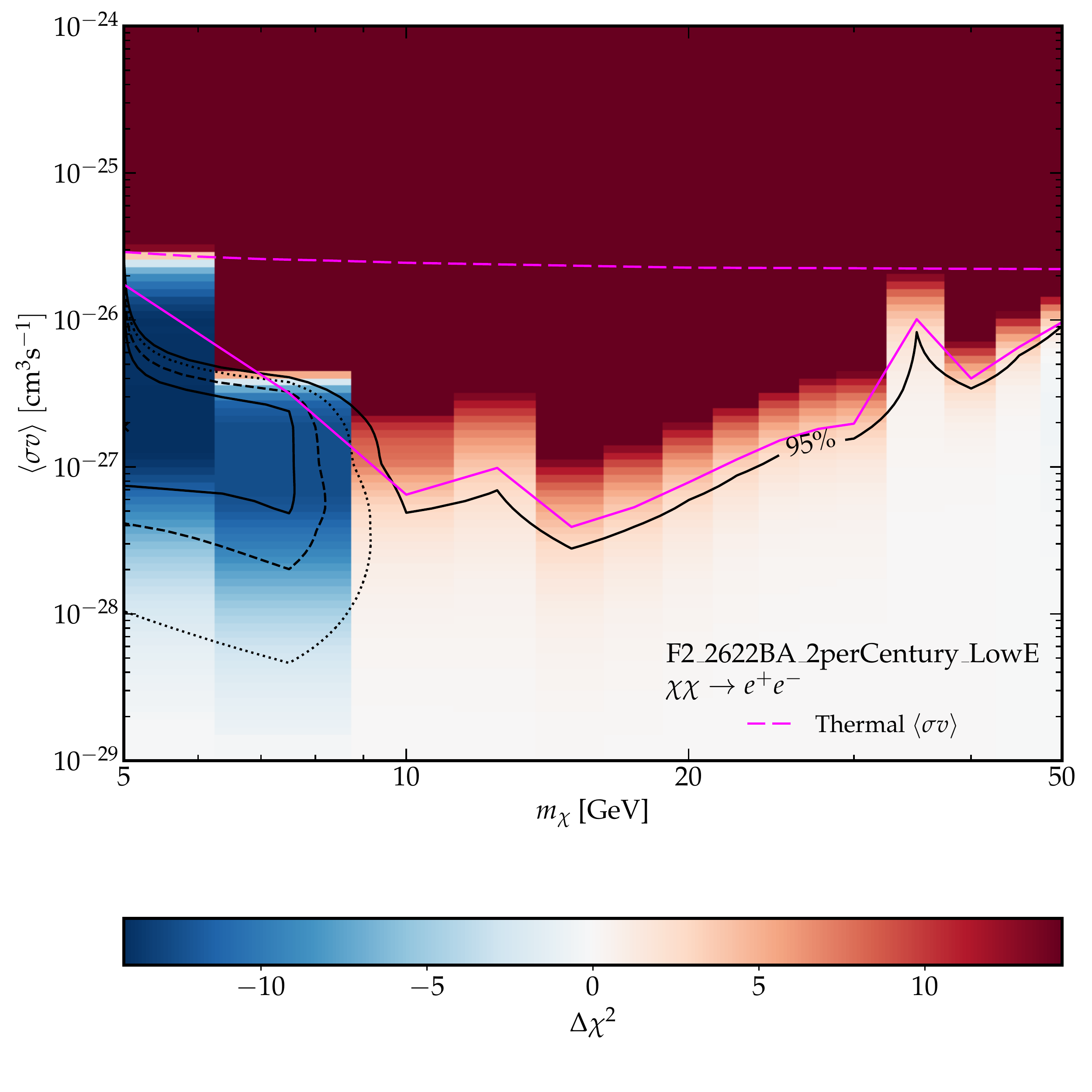}
\includegraphics[width=3.50in,angle=0]{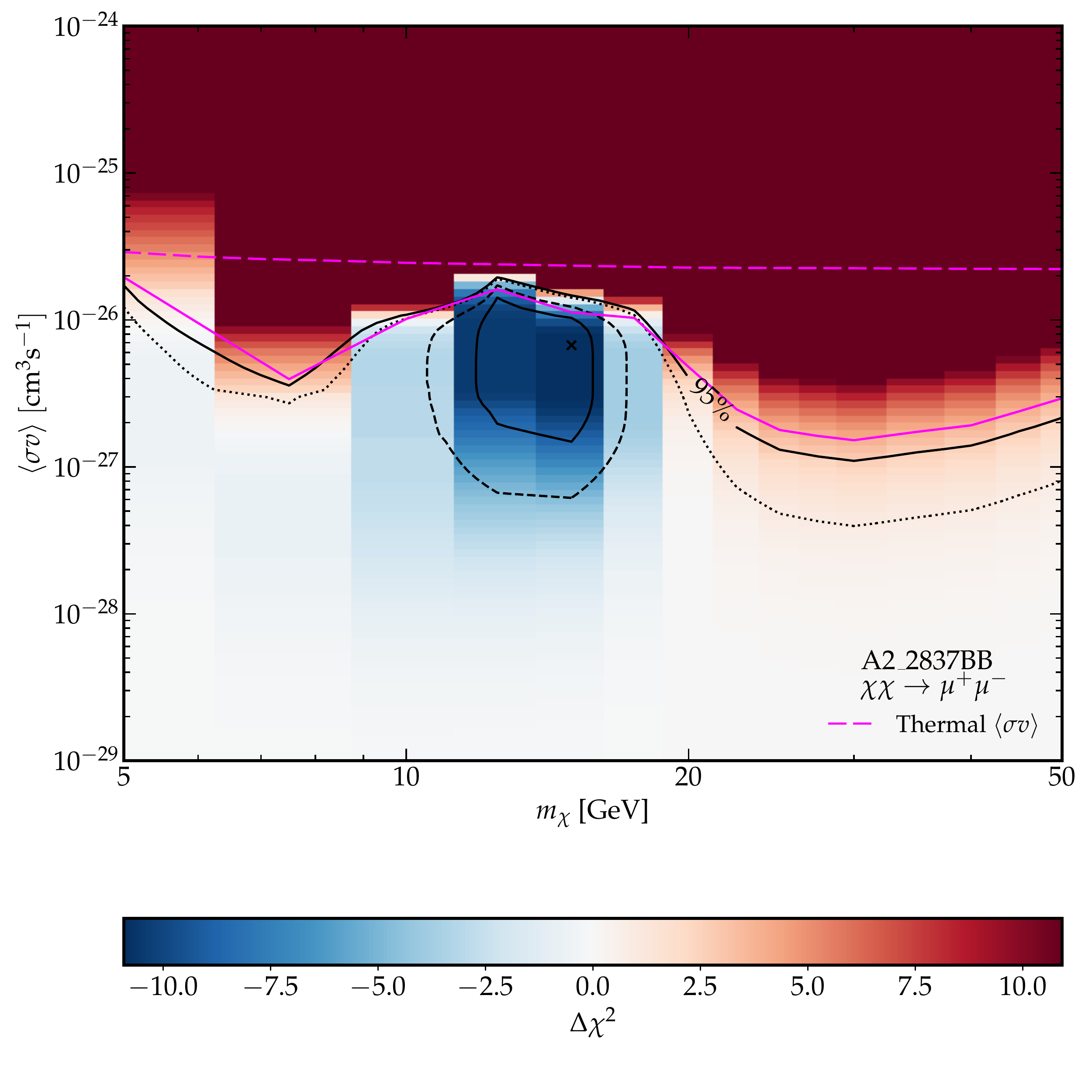}
\end{centering}
\vspace{-0.5cm}
\vspace{-0.0cm}

\begin{centering}
\hspace{-0.1cm}
\includegraphics[width=3.50in,angle=0]{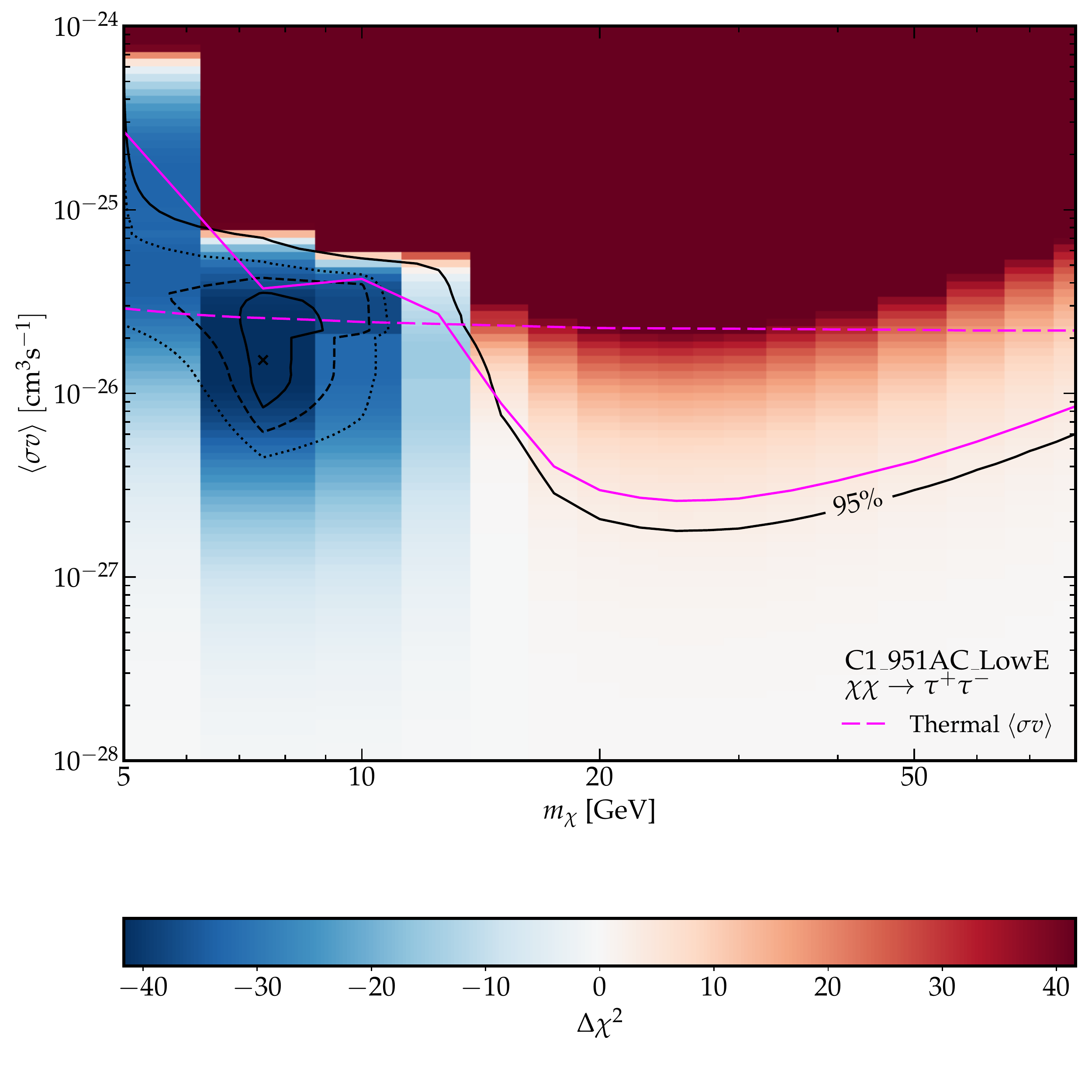}
\includegraphics[width=3.50in,angle=0]{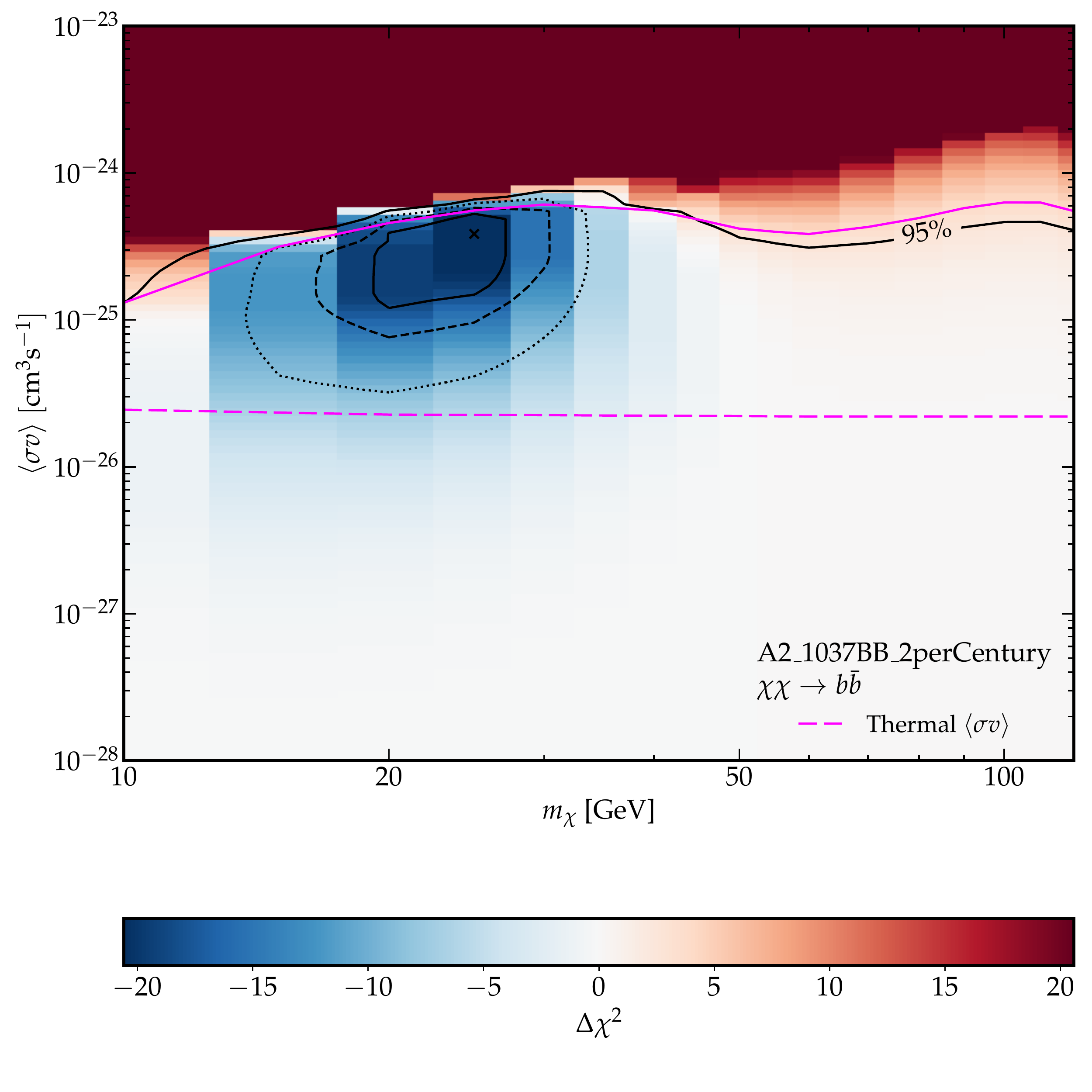}
\end{centering}
\vspace{-0.5cm}
\caption{Similar to Fig.~\ref{fig:heatmaps}, where for the same one background per channel, we show in solid magenta lines the upper limits derived by our second statistical method.}
\vspace{-0.0cm}
\label{fig:heatmaps_v2}
\end{figure*}

In this appendix, we explore a different statistical method for setting upper limits to the 
dark matter annihilation cross section. In our main analysis, we used a test statistic that tests 
the presence of a dark matter signal against the null hypothesis of purely conventional astrophysical 
sources to set our upper limits, i.e. $\langle \sigma v \rangle = 0$, which lies at the boundary of 
our parameter space. 
For this, we made use of Wilks' \cite{10.1214/aoms/1177732360} and Chernoff’s \cite{10.1214/aoms/1177728725} 
theorems. We will now discuss a slightly different approach that as we show gives similar upper limits. 
Our combined upper limit bands are roughly the same between the two methods.

Alternatively to the main text, we can use the test statistic,
\begin{equation}
    LR (\langle \sigma v \rangle) = 
- 2\log{\frac{\mathcal{L}(\langle \sigma v \rangle,\vec{\theta}_{\mathrm{max}})}{
    \mathcal{L}(\hat{\langle \sigma v \rangle},\hat{\vec{\theta}})}},
\end{equation}
which is equal to a $\chi^2$ difference.
With $\vec{\theta}$ we denote all the nuisance parameters of the astrophysical background, 
while with hats we denote the parameters that maximize the likelihood function.
According to Wilks' theorem, this test statistic follows a $\chi^2$-distribution with 1 degree of 
freedom since now we do not have a hypothesis that lies on the boundary of our parameter space.

Following the standard convention in literature, we can deduce a 95\% upper limit on the parameter 
$\langle \sigma v \rangle$, by the following procedure. For a fixed annihilation channel and dark 
matter mass, we find the parameters $\hat{\langle \sigma v \rangle}$ and $\hat{\vec{\theta}}$ that 
maximize the likelihood function, and then increase $\langle \sigma v \rangle$ until the test 
statistic $LR$ reaches the value 3.84. Within that procedure, $\vec{\theta}_{\mathrm{max}}$ are the values of the astrophysical nuisance parameters 
that maximize the likelihood for a given value of $\langle \sigma v \rangle$.
This method of profile likelihood is also called the MINOS method in high energy physics.

The upper limits with this method appear to be similar to the ones created with the method of our main analysis for almost every background.
For that reason we include Fig.~\ref{fig:heatmaps_v2}, where we also show the upper limits with this method for the backgrounds of Fig.~\ref{fig:heatmaps}.
From this figure, we can see that the upper limits set by this slightly different statistical methodology that doesn't rely on Chernoff's theorem are similar to the ones of our main analysis.
That is true for every background that we use to construct our upper limit bands.
Therefore we claim that the final combined bands of Fig.~\ref{fig:combined_bands} are robust to both 
astrophysical background uncertainties and also to the statistical treatment of the positron fraction 
measurement.

\bibliography{DarkMatter_Limits}

\end{document}